# Quantum Mechanics / Coarse-Grained Molecular Mechanics (QM/CG-MM)

Anton V. Sinitskiy and Gregory A. Voth[1, a)]

[1]Department of Chemistry, James Franck Institute, and Institute for Biophysical Dynamics, The University of Chicago, Chicago, Illinois 60637, United States

**ABSTRACT**

Numerous molecular systems, including solutions, proteins, and composite materials, can be modeled using mixed-resolution representations, of which the quantum mechanics/molecular mechanics (QM/MM) approach has become the most widely used. However, the QM/MM approach often faces a number of challenges, including the slow sampling of the large configuration space for the MM part, the high cost of repetitive QM computations for changing coordinates of atoms in the MM surroundings, and a difficulty in providing a simple, qualitative interpretation of numerical results in terms of the influence of the molecular environment upon the active QM region. In this paper, we address these issues by combining QM/MM modeling with the methodology of "bottom-up" coarse-graining (CG) to provide the theoretical basis for a systematic quantum-mechanical/coarse-grained molecular mechanics (QM/CG-MM) mixed resolution approach. A derivation of the method is presented based on a combination of statistical mechanics and quantum mechanics, leading to an equation for the effective Hamiltonian of the QM part, a central concept in the QM/CG-MM theory. A detailed analysis of different contributions to the effective Hamiltonian from electrostatic, induction, dispersion and exchange interactions between the QM part and the surroundings is provided, serving as a foundation for a potential hierarchy of QM/CG-MM methods varying in their accuracy and computational cost. A relationship of the QM/CG-MM methodology to other mixed resolution approaches is also discussed.

[a)]Author to whom correspondence should be addressed: gavoth@uchicago.edu

I.  INTRODUCTION

Mixed resolution methods are important to allow large heterogeneous systems to be efficiently studied at a reasonable computational cost. Such methods separate a molecular system under investigation into at least two parts, such that the portion(s) of the system that require high accuracy are treated with a computationally more expensive and more precise method, while the other portion(s) are treated at a lower, though still appropriate, level of description.

Among mixed-resolution methods, the quantum mechanics/molecular mechanics (QM/MM) approach[1] is arguably the most widely used one.[2-7] However, in many cases QM/MM methods prove to be suboptimal. If a system under investigation has a complex dynamics, with characteristic time scales of conformational changes spanning over multiple orders of magnitude (e.g., as in many enzymes or biomolecular complexes), QM/MM models may not provide sufficient sampling of the configuration space (both the QM and MM parts). Moreover, if diverse and important atomistic configurations of the surroundings are represented with comparable weights in the thermodynamic ensemble, the most expensive part of QM/MM computations, namely the QM one, needs to be repeated for numerous configurations of the surroundings, thereby greatly increasing the computational cost of the QM/MM model. In addition, one may encounter some difficulty in providing a simple, qualitative interpretation of numerical results in terms of influence of the molecular environment upon the active QM region, since the surroundings is characterized by a large number of MM degrees of freedom and their configurations.

On the other hand, coarse-grained (CG) modeling and simulation is another well-known, widely used, and computationally efficient approach that successfully addresses the problems similar to those listed above for the MM region.[8-13] By reducing the number of degrees of freedom used to describe a system, CG simulation can decrease the cost of computation by several orders of magnitude. As such, it allows for numerical simulations over larger length and time scales than do the atomistically resolved MM simulation models. In multiscale models, CG variables can also provide good candidates for collective variables to be used in enhanced sampling techniques at the atomistic level. The use of CG models therefore simplifies analysis of complex systems by focusing on their most important characteristics.

In this paper, we combine key ideas of QM/MM and CG modeling into a united "QM/CG-MM" approach and provide its fundamental theoretical underpinnings. Subsequent papers will develop and apply specific numerical algorithms associated with the presented theoretical ideas, hopefully in collaboration with experienced and interested electronic structure theorists. We develop the theory herein by adhering to the formal "bottom-up" statistical mechanics of the multiscale coarse-graining (MS-CG) theory[14-17] and related approaches such as relative entropy minimization.[18,19] Our QM/CG-MM approach should lead to a synergistic increase in efficiency of simulations due to the following three factors: First, it provides tools for enhanced sampling of the surrounding configurations. Second, it leads to a reduction in the required amount of expensive QM computations by averaging the influence of the surroundings on the active part. Third, the methodology should simplify the analysis of the effect of the surroundings on the QM part. It does so by clustering together similar fine-grained atomistic configurations and replacing a dependence on numerous atomistic degrees of freedom of the effective energies and wave functions of the QM part by their dependence on a smaller number of CG variables.

We note that some efforts to formulate QM calculations with a CG (or mixed resolution MM and CG) environment have previously appeared in the literature.[20-24] For the most part, these methods mainly blend together the QM and CG pieces, rather than utilizing the statistical mechanics of systematic bottom-up coarse-graining inherent, e.g., in the MS-CG approach. In the present paper, we instead develop then latter approach, formalizing the concept of QM/CG-MM modeling and creating a basis for both its implementations and for various approximations to the effective QM Hamiltonian in the QM/CG-MM model.

## II. DEFINITONS OF MODELS

This section formulates two representations of a molecular system to be modeled. One representation is considered here as exact (though in the present version it relies on the Born-Oppenheimer approximation), while the other formalizes the idea of the QM/CG-MM approach. Consistency between the two representations serves as a basis for further derivations. From now on, the term "the system" will be used for the total of the active part described at the quantum resolution level (the "QM" part) and the spectator part is to be coarse-grained (the "surroundings").

### A. Full description of the system

In the Born-Oppenheimer approximation, the exact partition function $Z$ of the system at a given inverse temperature $\beta = (k_B T)^{-1}$ (assuming classical nuclei) can be written as

$$Z = \int \frac{d\mathbf{r}_n d\mathbf{p}_n}{(2\pi\hbar)^{3N_n^{sys}}} \mathrm{Tr}_{r_e}\left\{e^{-\beta \hat{H}}\right\} = \int \frac{d\mathbf{r}_n d\mathbf{p}_n}{(2\pi\hbar)^{3N_n^{sys}}} e^{-\beta T_n(\mathbf{p}_n)} \mathrm{Tr}_{r_e}\left\{e^{-\beta \hat{H}_e}\right\}, \qquad (1)$$

where $\mathbf{r}_n$ and $\mathbf{p}_n$ are the coordinates and momenta of the nuclei in the system, respectively, $N_n^{sys}$ is the total number of the nuclei in the system, the trace is taken over the positions of all the electrons in the system $\mathbf{r}_e$, $\hat{H}$ is the full Hamiltonian operator of the system, $T_n$ is the classical kinetic energy of the nuclei in the system, and $\hat{H}_e$ is the electronic Hamiltonian. Then, the expectation value $\langle \hat{O} \rangle$ of a quantum mechanical operator $\hat{O}$ at equilibrium can be expressed as

$$\langle \hat{O} \rangle = \frac{1}{Z} \int \frac{d\mathbf{r}_n d\mathbf{p}_n}{(2\pi\hbar)^{3N_n^{sys}}} e^{-\beta T_n(\mathbf{p}_n)} \mathrm{Tr}_{r_e} \left\{ e^{-\beta \hat{H}_e} \hat{O} \right\}. \quad (2)$$

Averaging in Eq. (2) is fulfilled in both the quantum and thermodynamic sense: by surrounding the operator $\hat{O}$ with bra and ket vectors (implied in the trace), and by summation over different states with the temperature-dependent Boltzmann factor, respectively. Note that the operators $\hat{H}$ and $\hat{H}_e$ act in a $N_e^{sys}$–particle Hilbert space, where $N_e^{sys}$ is the total number of the electrons in the system.

## B. QM/CG-MM representation

As explained in the Introduction, one of the goals of the present QM/CG-MM approach is to express the influence of the surroundings on the QM part in terms of CG variables describing the surroundings. A strict approach to this problem, accounting for the charge transfer between the QM part and the surroundings, is outlined in Appendix A. However, in most cases it will likely make sense to choose the boundary between the QM and CG-MM parts of the system in such a way that charge transfer across this boundary is negligible and so the number of electrons in the QM part is therefore a well-defined integer number.

Under the assumption that the boundary is chosen in this way, we split the set of the coordinates of all the electrons in the system $\{\mathbf{r}_e\}$ into the set of the coordinates of the electrons in the QM part $\{\mathbf{r}_e^{QM}\}$ and the set of the coordinates of the electrons in the surroundings $\{\mathbf{r}_e^{surr}\}$:

$$\{\mathbf{r}_e\} = \{\{\mathbf{r}_e^{QM}\}, \{\mathbf{r}_e^{surr}\}\}. \quad (3)$$

The resulting formulas for the physical observables should account for indistinguishability of all the electrons in the system. We also split the set of the coordinates of all the nuclei in the system $\{\mathbf{r}_n\}$ into the set of the coordinates of the nuclei in the QM part $\{\mathbf{r}_n^{QM}\}$ and the set of the coordinates of the nuclei in the surroundings $\{\mathbf{r}_n^{surr}\}$:

$$\{\mathbf{r}_n\} = \{\{\mathbf{r}_n^{QM}\}, \{\mathbf{r}_n^{surr}\}\}, \quad (4)$$

but this time we assume that all the nuclei are distinguishable, implying that temperature of the system is high enough so that the quantum statistics nature of the nuclei beyond Boltzmann statistics does not manifest itself.

Now, we introduce an effective Hamiltonian for the QM part $\hat{H}_{eff}^{QM}\left(\mathbf{r}_e^{QM}; \mathbf{r}_n^{QM}, \mathbf{R}^N\right)$ as an operator that acts only on the coordinates of the electrons in the QM part $\mathbf{r}_e^{QM}$, and also parametrically depends on the coordinates of the nuclei in the QM part $\mathbf{r}_n^{QM}$ and the CG coordinates $\mathbf{R}^N$ for the surroundings, but not on the coordinates of individual electrons or nuclei in the surroundings. The operator $\hat{H}_{eff}^{QM}$ acts in a $N_e^{QM}$–particle Hilbert space, where $N_e^{QM}$ is the number of the electrons in the QM part.

From the viewpoint of the QM/CG-MM approach, it is reasonable to define the expectation value $\langle\hat{O}\rangle_{QM/CG-MM}$ of an operator $\hat{O} = \hat{O}\left(\mathbf{r}_e^{QM}; \mathbf{r}_n^{QM}\right)$ with the use of the effective Hamiltonian $\hat{H}_{eff}^{QM}\left(\mathbf{r}_e^{QM}; \mathbf{r}_n^{QM}, \mathbf{R}^N\right)$ in the following way:

$$\langle\hat{O}\rangle_{QM/CG-MM} = \frac{1}{Z_{QM/CG-MM}} \int \frac{d\mathbf{r}_n^{QM} d\mathbf{p}_n^{QM}}{(2\pi\hbar)^{3N_n^{QM}}} \int \frac{d\mathbf{R}^N d\mathbf{P}^N}{(2\pi\hbar)^{3N}} e^{-\beta\left[T_n^{QM}(\mathbf{p}_n^{QM}) + T_{CG}(\mathbf{P}^N)\right]} \mathrm{Tr}_{\mathbf{r}_e^{QM}}\left\{e^{-\beta\hat{H}_{eff}^{QM}} \hat{O}\right\}, \quad (5)$$

where

$$Z_{QM/CG-MM} = \int \frac{d\mathbf{r}_n^{QM} d\mathbf{p}_n^{QM}}{(2\pi\hbar)^{3N_n^{QM}}} \int \frac{d\mathbf{R}^N d\mathbf{P}^N}{(2\pi\hbar)^{3N}} e^{-\beta\left[T_n^{QM}(\mathbf{p}_n^{QM}) + T_{CG}(\mathbf{P}^N)\right]} \text{Tr}_{r_e^{QM}} \left\{ e^{-\beta \hat{H}_{eff}^{QM}} \right\}. \qquad (6)$$

Here, $N_n^{QM}$ is the number of nuclei in the QM part, $N$ is the number of the CG particles in the surroundings, $T_n^{QM}$ is the kinetic energy of the nuclei in the QM part, $T_{CG}$ is the kinetic energy of the CG particles, $\mathbf{p}_n^{QM}$ and $\mathbf{P}^N$ are the momenta conjugate to $\mathbf{r}_n^{QM}$ and $\mathbf{R}^N$, respectively, and the operator $\hat{O} = \hat{O}(\mathbf{r}_e^{QM}; \mathbf{r}_n^{QM})$ acts only on the coordinates of the electrons in the QM part $\mathbf{r}_e^{QM}$, and may also parametrically depend on the coordinates of the nuclei in the QM part $\mathbf{r}_n^{QM}$. The traces in Eqs. (5) and (6) are taken only over the positions of the electrons in the QM part, while in Eqs. (1) and (2) the traces are taken over the positions of all the electrons in the system.

C. Consistency condition

To introduce an explicit expression for $\hat{H}_{eff}^{QM}$, we impose the following *consistency condition*: The expectation values of an arbitrary quantum mechanical operator at a given temperature predicted by the QM/CG-MM approach $\langle \hat{O} \rangle_{QM/CG-MM}$ must coincide with that predicted by the full quantum mechanical description of the system $\langle \hat{O} \rangle$:

$$\langle \hat{O} \rangle_{QM/CG-MM} = \langle \hat{O} \rangle. \qquad (7)$$

This consistency condition motivates the following definition of the effective QM Hamiltonian (for the derivation, see Sec. 1 of the Supplementary Material):

$$e^{-s\hat{H}_{eff}^{QM}} = C \cdot \int d\mathbf{r}_n^{surr} \mathsf{u}\left(\mathbf{R}^N - \mathbf{M}_\mathbf{R}^N(\mathbf{r}_n^{surr})\right) \text{Tr}_{r_e^{surr}} \left\{ e^{-s\hat{H}_e} \right\}, \qquad (8)$$

where $\mathbf{M}_\mathbf{R}^N(\mathbf{r}_n^{surr}) = \{\mathbf{M}_{\mathbf{R}_I}(\mathbf{r}_n^{surr})\}$, with $I = 1, \ldots, N$, is the set of the mapping operators that coarse-grain the coordinates of the nuclei in the surroundings $\mathbf{r}_n^{surr}$ into a smaller number of the CG variables $\mathbf{R}^N = \{\mathbf{R}_1, \ldots, \mathbf{R}_N\}$ (for a discussion of mapping operators in the context of CG, see ref. [16]), and the following shorthand identity is used:

$$\mathsf{u}\left(\mathbf{R}^N - \mathbf{M}_\mathbf{R}^N(\mathbf{r}_n^{surr})\right) \equiv \prod_{I=1}^{N} \mathsf{u}\left(\mathbf{R}_I - \mathbf{M}_{\mathbf{R}_I}(\mathbf{r}_n^{surr})\right) . \tag{9}$$

The trace on the right hand side in Eq. (8) is defined in the following way:

$$\mathrm{Tr}_{\mathbf{r}_e^{surr}}\left\{e^{-s\hat{H}_e}\right\} = \left(N_e^{QM}!\right)\left(N_e^{sys}!\right)\int d\mathbf{r}_e^{QM} d\mathbf{r}_e'^{QM} d\mathbf{r}_e^{surr} \left(\mathcal{A}_{QM}|\mathbf{r}_e^{QM}\rangle\right)$$
$$\times \langle \mathbf{r}_e^{QM}, \mathbf{r}_e^{surr}|\mathcal{A}_{sys} e^{-s\hat{H}_e}\mathcal{A}_{sys}|\mathbf{r}_e'^{QM}, \mathbf{r}_e^{surr}\rangle\left(\langle \mathbf{r}_e'^{QM}|\mathcal{A}_{QM}\right) \tag{10}$$

to account for indistinguishability of the electrons in the QM part and the surroundings (for details, see Sec. 1 of the Supplementary Material). In Eq. (10), $\mathcal{A}_{sys}$ is the antisymmetrization operator for all the electrons in the system, and $\mathcal{A}_{QM}$ is the antisymmetrization operator for the electrons in the QM part:

$$\mathcal{A}_{sys} = \frac{1}{N_e^{sys}!}\sum_{\hat{P} \in S_{sys}}(-1)^P \hat{P}, \quad \mathcal{A}_{QM} = \frac{1}{N_e^{QM}!}\sum_{\hat{P} \in S_{QM}}(-1)^P \hat{P}, \tag{11}$$

where $S_{sys}$ and $S_{QM}$ are the sets of all possible permutations of electrons in the system and in the QM part, respectively. Finally, the constant $C$ in Eq. (8) is defined in the following way that depends on the choice of the functional form of the CG kinetic energy $T_{CG}$:

$$C = \frac{1}{\left(N_e^{QM}!\right)^2} \frac{\left(\int \frac{d\mathbf{p}_n^{surr}}{(2f\hbar)^{3N_n^{surr}}} e^{-sT_n^{surr}(\mathbf{p}_n^{surr})}\right)}{\left(\int \frac{d\mathbf{P}^N}{(2f\hbar)^{3N}} e^{-sT_{CG}(\mathbf{P}^N)}\right)}. \tag{12}$$

A discussion of the issue of the localization of the operator $\hat{O}$ to the QM part, and the consequences of that for the definition of the effective QM Hamiltonian in the QM/CG-MM method, are provided in Appendix B.

## III. EFFECT OF THE QM PART ON THE SURROUNDINGS

The expectation value of a function $O_{surr}\left(\mathbf{M}_{\mathbf{R}}^{N}(\mathbf{r}_{n}^{surr})\right)$ that depends only on the coordinates of the nuclei in the surrounding $\{\mathbf{r}_{n}^{surr}\}$ and *does so only via the mapping operators* $\mathbf{M}_{\mathbf{R}}^{N}$ can be written as

$$\langle O_{surr} \rangle = \frac{1}{Z} \int \frac{d\mathbf{r}_n d\mathbf{p}_n}{(2f\hbar)^{3N_n^{sys}}} e^{-\beta T_n(\mathbf{p}_n)} \text{Tr}_{r_e} \left\{ e^{-\beta \hat{H}_e} \right\} O_{surr}\left(\mathbf{M}_{\mathbf{R}}^{N}(\mathbf{r}_{n}^{surr})\right). \tag{13}$$

As demonstrated in Sec. 2 of the Supplementary Material, this expectation value of $O_{surr}$ can be rewritten solely in terms of CG variables, specifically,

$$\langle O_{surr} \rangle = \int d\mathbf{R}^{N} p_{surr}\left(\mathbf{R}^{N}\right) O_{surr}\left(\mathbf{R}^{N}\right), \tag{14}$$

where $p_{surr}(\mathbf{R}^N)$ is the probability distribution for the CG variables. This result, Eq. (14), is analogous to the statement that can be proven in the MS-CG method[16] that the expectation value of a function dependent on fine-grained variables only via the mapping operators can be accurately computed using the CG probability distribution, without a need to resort to the fine-grained representation.

As proven in Sec. 2 of the Supplementary Material, the probability distribution $p_{surr}(\mathbf{R}^N)$ can be exactly expressed solely in terms of the eigenfunctions of the effective QM Hamiltonian $E_{eff,i}(\mathbf{r}_n^{QM}, \mathbf{R}^N)$, namely:

$$p_{surr}\left(\mathbf{R}^N\right) = \frac{\int d\mathbf{r}_n^{QM} \sum_i e^{-\varsigma E_{\text{eff},i}(\mathbf{r}_n^{QM}, \mathbf{R}^N)}}{\int d\mathbf{R}^N d\mathbf{r}_n^{QM} \sum_i e^{-\varsigma E_{\text{eff},i}(\mathbf{r}_n^{QM}, \mathbf{R}^N)}}. \tag{15}$$

Hence, the concept of the effective QM Hamiltonian $\hat{H}_{\text{eff}}^{QM}$ plays the central role in the QM/CG-MM method. Being introduced in Sec. II to deal with the QM part, the same effective Hamiltonian, as shown in this section, is also sufficient for the full statistical mechanical description of the surroundings (at the CG level of representation).

## IV. MATRIX REPRESENTATION OF THE EFFECTIVE HAMILTONIAN

For the purposes of subsequent analysis, it is convenient to derive an expression for the effective Hamiltonian in a matrix representation. First, we introduce the required notation. We denote the electronic Hamiltonian for the QM part in the absence of the surroundings (i.e., in vacuum) as $\hat{H}_e^{QM}$, the electronic Hamiltonian for the surroundings in the absence of the QM part as $\hat{H}_e^{surr}$, and the potential energy of interaction between the QM part and the surroundings as $V_{QM-surr}$. Then the full electronic Hamiltonian of the system included into Eq. (1) can be written as

$$\hat{H}_e = \hat{H}_e^{QM}\left(\mathbf{r}_e^{QM}; \mathbf{r}_n^{QM}\right) + \hat{H}_e^{surr}\left(\mathbf{r}_e^{surr}; \mathbf{r}_n^{surr}\right) + V_{QM-surr}\left(\mathbf{r}_e^{QM}, \mathbf{r}_n^{QM}, \mathbf{r}_e^{surr}, \mathbf{r}_n^{surr}\right). \tag{16}$$

Further, we denote the eigenfunctions and the eigenvalues of $\hat{H}_e^{QM}$ within the Born-Oppenheimer approximation as $\Psi_{n_{QM}}^{QM}\left(\mathbf{r}_e^{QM}; \mathbf{r}_n^{QM}\right)$ and $E_{n_{QM}}^{QM}\left(\mathbf{r}_n^{QM}\right)$, respectively, and those of $\hat{H}_e^{surr}$ as $\Psi_{n_{surr}}^{surr}\left(\mathbf{r}_e^{surr}; \mathbf{r}_n^{surr}\right)$ and $E_{n_{surr}}^{surr}\left(\mathbf{r}_n^{surr}\right)$, respectively. Indices $n_{QM}$ and $n_{surr}$ here enumerate the eigenstates. Un-normalized wave functions and the spectrum of the system can be written as:

$$\Psi^{sys}_{n_{QM},n_{surr}} = \mathcal{A}_{sys}\left(\Psi^{QM}_{n_{QM}}\Psi^{surr}_{n_{surr}} + \sum_{k,l} c^{k,l}_{n_{QM},n_{surr}}\Psi^{QM}_{k}\Psi^{surr}_{l}\right), \quad (17)$$

$$E^{sys}_{n_{QM},n_{surr}} = E^{QM}_{n_{QM}}\left(\mathbf{r}^{QM}_{n}\right) + E^{surr}_{n_{surr}}\left(\mathbf{r}^{surr}_{n}\right) + \Delta E_{n_{QM},n_{surr}}\left(\mathbf{r}^{QM}_{n},\mathbf{r}^{surr}_{n}\right), \quad (18)$$

where $c^{n_{QM},n_{surr}}_{k,l}$ are coefficients and $\Delta E_{n_{QM},n_{surr}}$ is the energy of interaction between the QM part and the surroundings. We will further refer to $\Delta E_{n_{QM},n_{surr}}\left(\mathbf{r}^{QM}_{n},\mathbf{r}^{surr}_{n}\right)$ as the fine-grained interaction energy between the QM part and the surroundings because it depends on the fine-grained degrees of freedom for the surroundings $\mathbf{r}^{surr}_{n}$ and $n_{surr}$. In the perturbative regime, every eigenstate of the system can be associated with a specific electronic state of the QM part and a specific electronic state of the surroundings, which motivates the enumeration of the eigenstates of the system in Eqs. (17) and (18) by the pair of indices ($n_{QM}$, $n_{surr}$), referring to the electronic state of the QM part and the surroundings, respectively. However, Eqs. (17) and (18) are exact in a general case too, since the eigenfunctions of the Hamiltonians $\hat{H}^{QM}_{e}$ and $\hat{H}^{surr}_{e}$ form complete basis sets, and $\Delta E_{n_{QM},n_{surr}}$ is in essence defined by Eq. (18).

In the introduced notation, the matrix elements of $e^{-s\hat{H}^{QM}_{eff}}$ in the basis of the unperturbed wave functions of the QM part can be written in the following form (see Sec. 3 of the Supplementary Material for the derivation):

$$\begin{aligned}\langle i^{(0)}|e^{-s\hat{H}^{QM}_{eff}}|j^{(0)}\rangle &= \mathsf{u}_{i,j}e^{-sV^{surr}_{CG}(\mathbf{R}^{N})}e^{-sE^{QM}_{i}(\mathbf{r}^{QM}_{n})}\langle e^{-s\Delta E_{i,n_{surr}}(\mathbf{r}^{QM}_{n},\mathbf{r}^{surr}_{n})}\rangle_{\mathbf{R}^{N}} + e^{-sV^{surr}_{CG}(\mathbf{R}^{N})}\\ &\times\sum_{n_{QM}}e^{-sE^{QM}_{n_{QM}}(\mathbf{r}^{QM}_{n})}\langle k^{n_{QM},n_{surr}}_{i,j}(\mathbf{r}^{QM}_{n},\mathbf{r}^{surr}_{n})e^{-s\Delta E_{n_{QM},n_{surr}}(\mathbf{r}^{QM}_{n},\mathbf{r}^{surr}_{n})}\rangle_{\mathbf{R}^{N}}.\end{aligned} \quad (19)$$

where $|j^{(0)}\rangle$ is a short-hand notation for $\Psi^{QM}_{j}$. In Eq. (19), $V^{surr}_{CG}(\mathbf{R}^{N})$ is the CG potential for the surroundings in the absence of the QM part, defined as

$$V_{CG}^{surr}(\mathbf{R}^N) = -\frac{1}{S}\ln\left[C\sum_{n_{surr}}\int d\mathbf{r}_n^{surr}\mathsf{u}\left(\mathbf{R}^N - \mathbf{M}_\mathbf{R}^N(\mathbf{r}_n^{surr})\right)e^{-S E_{n_{surr}}^{surr}(\mathbf{r}_n^{surr})}\right], \tag{20}$$

with $C$ given by Eq. (12). This potential is analogous to the CG potential previously introduced in the MS-CG theory,[16] but differs from it in that $V_{CG}^{surr}(\mathbf{R}^N)$ allows for electronic excitations (as implied by the summation over $n_{surr}$) in the CG (sub)system (in this case, the surroundings) The averaging $\langle \cdots \rangle_{\mathbf{R}^N}$ in Eq. (19) is defined by

$$\langle f \rangle_{\mathbf{R}^N} = \frac{\sum_{n_{surr}}\int d\mathbf{r}_n^{surr}\mathsf{u}\left(\mathbf{R}^N - \mathbf{M}_\mathbf{R}^N(\mathbf{r}_n^{surr})\right)e^{-S E_{n_{surr}}^{surr}(\mathbf{r}_n^{surr})}f(\mathbf{r}_n^{surr},n_{surr})}{\sum_{n_{surr}}\int d\mathbf{r}_n^{surr}\mathsf{u}\left(\mathbf{R}^N - \mathbf{M}_\mathbf{R}^N(\mathbf{r}_n^{surr})\right)e^{-S E_{n_{surr}}^{surr}(\mathbf{r}_n^{surr})}}, \tag{21}$$

and practical aspects of its computation are discussed in Appendix C. Note that this averaging is performed not only over various conformations of the surroundings, as implied by the integration over $\mathbf{r}_n^{surr}$, but also over various electronic states in the surroundings, due to the summation over $n_{surr}$. The coefficients $k_{i,j}^{n_{QM},n_{surr}}$ in Eq. (19) are defined by

$$k_{i,j}^{n_{QM},n_{surr}} = \frac{\langle i^{(0)}|\mathrm{Tr}_{r_e^{surr}}\{|n_{QM}n_{surr}\rangle\langle n_{QM}n_{surr}|\}|j^{(0)}\rangle}{\langle n_{QM}n_{surr}|n_{QM}n_{surr}\rangle} - \mathsf{u}_{i,n_{QM}}\mathsf{u}_{j,n_{QM}}, \tag{22}$$

where $|n_{QM}n_{surr}\rangle$ is a short-hand notation for $\Psi_{n_{QM},n_{surr}}^{sys}$. From the viewpoint of perturbation theory, the coefficients $k_{i,j}^{n_{QM},n_{surr}}$ are different from zero only due to exchange interactions and polarization of the QM part and the surroundings by each other. Hence, in the cases of weak interaction between the QM part and the surroundings the terms $k_{i,j}^{n_{QM},n_{surr}}$ may be considered as small parameters.

The computation of the matrix $\langle i^{(0)} | e^{-s\hat{H}_{eff}^{QM}} | j^{(0)} \rangle$ by Eq. (19), with the subsequent diagonalization of this matrix, yields a general solution to the problem of finding the effective QM Hamiltonian $\hat{H}_{eff}^{QM}$. Indeed, the eigenfunctions of $\hat{H}_{eff}^{QM}$ coincide with those of $e^{-s\hat{H}_{eff}^{QM}}$, while the eigenvalues $E_{eff,i}$ of the former operator are related to the eigenfunctions $\lambda_i$ of the latter as $E_{eff,i} = -\ln \lambda_i / s$. Then, the sought-after effective QM Hamiltonian can be written in terms of its eigenvalues and eigenfunctions via a spectral representation.

## V. ANALYSIS OF CONTRIBUTIONS TO THE EFFECTIVE HAMILTONIAN

### A. Electrostatic interactions between the QM part and the surroundings

By analogy with an analysis of intermolecular interactions,[25] we first consider the approximation that no exchange interaction between the two parts takes place and that the wave functions of the QM part and the surroundings do not differ from those in the two isolated parts of the system. In this case, all coefficients $k_{i,j}^{n_{QM},n_{surr}}$ equal zero and, as follows from Eq. (19),

$$\langle i^{(0)} | e^{-s\hat{H}_{eff}^{QM}} | j^{(0)} \rangle = \delta_{i,j} e^{-s E_{eff,i}^{(elstat)}}, \qquad (23)$$

where

$$E_{eff,n_{QM}}^{(elstat)} = E_{n_{QM}}^{QM}\left(\mathbf{r}_n^{QM}\right) + V_{CG}^{surr}\left(\mathbf{R}^N\right) + \Delta E_{eff,n_{QM}}^{(elstat)}\left(\mathbf{r}_n^{QM}, \mathbf{R}^N\right) \qquad (24)$$

with $E_{n_{QM}}^{QM}\left(\mathbf{r}_n^{QM}\right)$ being the energy of the isolated QM part in quantum state $n_{QM}$ as defined in Sec. IV, $V_{CG}^{surr}\left(\mathbf{R}^N\right)$ being the CG potential for the isolated surroundings as defined by Eq. (20), and $\Delta E_{eff,n_{QM}}^{(elstat)}$ is the coarse-grained effective energy of interactions between the QM part in quantum state $n_{QM}$ and the surroundings, defined here as

$$\Delta E_{eff,n_{QM}}^{(\text{elstat})}\left(\mathbf{r}_n^{QM},\mathbf{R}^N\right) = -\frac{1}{S}\ln\langle e^{-S\Delta E_{n_{QM},n_{surr}}^{(\text{elstat})}(\mathbf{r}_n^{QM},\mathbf{r}_n^{surr})}\rangle_{\mathbf{R}^N}, \tag{25}$$

that is, by averaging the fine-grained energy of interactions between the QM part and the surroundings defined in Sec. IV, Eq. (18). Indices $i$ and $j$ in Eq. (23) and $n_{QM}$ in Eqs. (24) and (25) enumerate the electronic states in the QM part, as defined above in Sec. IV.

Hence, in the approximation of only electrostatic interaction between the two parts of the system, *which is analogous to the mechanical embedding in QM/MM schemes*,[2,4,26] the eigenfunctions $|i\rangle$ of the effective QM Hamiltonian $\hat{H}_{eff}^{QM}$ coincide with those of the electronic Hamiltonian of the isolated QM part $\hat{H}_e^{QM}$:

$$|i\rangle = |i^{(0)}\rangle, \tag{26}$$

while the effective energy levels of the QM part $E_{eff,n_{QM}}^{(\text{elstat})}$ are sums of the corresponding energy of the QM part in vacuum, the free energy of the CG surroundings, and the electrostatic contribution to the effective energy $\Delta E_{eff,n_{QM}}^{(\text{elstat})}$. Equation (25) most succinctly reflects the key difference between the systematic bottom-up approach presented in this paper and blending together the QM and CG pieces in previous publications:[20-24] The effective potential of interactions between the two subsystems is obtained by averaging the exponentials of the fine-grained interaction energy between the two subsystems in the units of $k_BT$, $e^{-S\Delta E_{n_{QM},n_{surr}}^{(\text{elstat})}}$, and by taking the logarithm of the average, rather than directly averaging the fine-grained interaction energy $\Delta E_{n_{QM},n_{surr}}^{(elstat)}$.

Then, the expectation value of a quantum mechanical operator $\hat{O}$ can be computed as

$$\langle \hat{O} \rangle_{QM/CG-MM} = \frac{\int d\mathbf{r}_n^{QM} d\mathbf{R}^N \sum_{n_{QM}} e^{-\varsigma E_{eff,n_{QM}}^{(elstat)}(\mathbf{r}_n^{QM}, \mathbf{R}^N)} \langle n_{QM}^{(0)} | \hat{O} | n_{QM}^{(0)} \rangle}{\int d\mathbf{r}_n^{QM} d\mathbf{R}^N \sum_{n_{QM}} e^{-\varsigma E_{eff,n_{QM}}^{(elstat)}(\mathbf{r}_n^{QM}, \mathbf{R}^N)}}, \quad (27)$$

and the partition function of the system as

$$Z = \left( \int \frac{d\mathbf{p}_n^{QM} e^{-\varsigma T_n^{QM}(\mathbf{p}_n^{QM})}}{(2f\hbar)^{3N_n^{QM}}} \right) \left( \int \frac{d\mathbf{P}^N e^{-\varsigma T_{CG}(\mathbf{P}^N)}}{(2f\hbar)^{3N}} \right) \sum_{n_{QM}} \int d\mathbf{r}_n^{QM} d\mathbf{R}^N e^{-\varsigma E_{eff,n_{QM}}^{(elstat)}(\mathbf{r}_n^{QM}, \mathbf{R}^N)}. \quad (28)$$

with the contributions to the eigenvalues of the effective Hamiltonian restricted to the purely electrostatic term, $E_{eff,n_{QM}} = E_{eff,n_{QM}}^{(elstat)}$.

We consider now the computation of $E_{eff,n_{QM}}^{(elstat)}$ required in Eqs. (27) and (28) in more detail. The first term in Eq. (24) is the $n_{QM}$-th energy level of the isolated QM part that remains the same for different values of the CG variables $\mathbf{R}^N$, but depends on $\mathbf{r}_n^{QM}$. The second term is the free energy of the surrounding media (e.g., solvent) that can be computed separately based, e.g., on the MS-CG method[15,16] or by other means without the need to perform expensive QM computations. Possible strategies to compute the third term $\Delta E_{eff,n_{QM}}^{(elstat)}$ are discussed in Appendix D, and additional technical aspects are considered in Sec. 4 of the Supplementary Material. An outline of an algorithm summarizing the discussion in this subsection is presented in Scheme 1 of Appendix E.

### B. Induction interactions between the QM part and the surroundings

Continuing the analogy with a standard analysis of intermolecular interactions, we now account for polarization of the QM part by the unperturbed surroundings and that of the surroundings by the unperturbed QM part. We still assume that there is no exchange interaction between the QM part and the surroundings, and we neglect for now dispersion interactions between the two parts.

The exact expressions for $k_{i,j}^{n_{QM},n_{surr}}$ in terms of the coefficients $c_{n_{QM},n_{surr}}^{k,l}$ are given in Sec. 5 of the Supplementary Material. To proceed with the computations of eigenvalues and eigenfunctions of the effective QM Hamiltonian, we need explicit expressions for the coefficients $c_{n_{QM},n_{surr}}^{k,n_{surr}}$, $c_{n_{QM},n_{surr}}^{n_{QM},l}$ and the fine-grained interaction energy $\Delta E_{n_{QM},n_{surr}}$. A simple way to get them is via perturbation theory, as discussed in Sec. 6 of the Supplemental Material. For analytical and illustrative purposes, below we consider the lowest-order nontrivial terms in these expansions (higher-order terms can be added if greater numerical accuracy of the computations is required):

$$c_{n_{QM},n_{surr}}^{k,n_{surr}} = -\frac{\langle k^{(0)} n_{surr}^{(0)} | V_{QM-surr} | n_{QM}^{(0)} n_{surr}^{(0)} \rangle}{E_k^{QM} - E_{n_{QM}}^{QM}} + O(v^2), \tag{29}$$

$$c_{n_{QM},n_{surr}}^{n_{QM},l} = -\frac{\langle n_{QM}^{(0)} l^{(0)} | V_{QM-surr} | n_{QM}^{(0)} n_{surr}^{(0)} \rangle}{E_l^{surr} - E_{n_{surr}}^{surr}} + O(v^2), \tag{30}$$

$$\Delta E_{n_{QM},n_{surr}}^{(\text{elstat+ind})} = \Delta E_{n_{QM},n_{surr}}^{(\text{elstat})} - \sum_{k \neq n_{QM}} \frac{|\langle k^{(0)} n_{surr}^{(0)} | V_{QM-surr} | n_{QM}^{(0)} n_{surr}^{(0)} \rangle|^2}{E_k^{QM} - E_{n_{QM}}^{QM}}$$
$$- \sum_{l \neq n_{surr}} \frac{|\langle n_{QM}^{(0)} l^{(0)} | V_{QM-surr} | n_{QM}^{(0)} n_{surr}^{(0)} \rangle|^2}{E_l^{surr} - E_{n_{surr}}^{surr}} + O(v^3). \tag{31}$$

where $\Delta E_{n_{QM},n_{surr}}^{(\text{elstat})}$ is the purely electrostatic contribution to the fine-grained interaction energy between the QM and CG parts discussed above in Sec. V.A, and $v$ is a small dimensionless parameter of expansion in these series, defined as

$$v = \frac{\langle V_{QM-surr} \rangle}{\langle \Delta E_{excit} \rangle}. \tag{32}$$

Here, $\langle V_{QM-surr} \rangle$ is a typical value of the matrix elements of the potential energy of interaction between the QM part and the surroundings $\langle k^{(0)} l^{(0)} | V_{QM-surr} | n_{QM}^{(0)} n_{surr}^{(0)} \rangle$ and $\langle \Delta E_{excit} \rangle$ is a typical value of the energy of electronic excitations in the QM part or the surroundings.

Computation of the terms to all orders in $v$ on the right hand side of Eq. (29), as well as some terms on the right hand side of Eq. (31), including the first and the second ones, can be performed in a way similar to that discussed in Sec. 4 of the Supplemental Material. Physically, these terms correspond to polarization of the QM part by the unperturbed surroundings.

By contrast, the terms of all order in $v$ on the right hand side of Eq. (30), as well as some terms on the right hand side of Eq. (31), including the third one, involve off-diagonal elements of $V_{QM\text{-}surr}$ over two different electronic states of the surroundings. Physically, these terms account for polarization of the surroundings by the unperturbed QM part. A straightforward computation of such terms demands the knowledge of the wave functions of the surroundings. However, the QM/CG-MM approach must deal with such terms without resorting to a computationally expensive quantum mechanical description of the surroundings. A possible strategy for resolving this problem is presented in Sec. 7 of the Supplemental Material. The final results are as follows:

$$c_{n_{QM},n_{surr}}^{k,n_{surr}} = -\frac{\langle k^{(0)} | V_{ext}^{elstat,n_{surr}} | n_{QM}^{(0)} \rangle}{E_k^{QM} - E_{n_{QM}}^{QM}} + O(v^2), \tag{33}$$

$$\sum_{l \neq n_{surr}} \left( c_{n_{QM},n_{surr}}^{n_{QM},l} \right)^2 = \frac{1}{2} \sum_{I \in CG} \sum_{\substack{r=x,y,z \\ s=x,y,z}} F_r^{QM,n_{QM}}(\mathbf{R}_I) s_{rs}^{surr,I,n_{surr}} F_s^{QM,n_{QM}}(\mathbf{R}_I) + O(v^3), \tag{34}$$

$$\begin{aligned}
\Delta E_{n_{QM},n_{surr}}^{(elstat+ind)} = \Delta E_{n_{QM},n_{surr}}^{(elstat)} &- \sum_{k \neq n_{QM}} \frac{\left| \langle k^{(0)} | V_{ext}^{elstat,n_{surr}} | n_{QM}^{(0)} \rangle \right|^2}{E_k^{QM} - E_{n_{QM}}^{QM}} \\
&- \frac{1}{2} \sum_{I \in CG} \sum_{\substack{r=x,y,z \\ s=x,y,z}} F_r^{QM,n_{QM}}(\mathbf{R}_I) r_{rs}^{surr,I,n_{surr}} F_s^{QM,n_{QM}}(\mathbf{R}_I) + O(v^3),
\end{aligned} \tag{35}$$

where $V_{ext}^{elstat,n_{surr}}$, defined as

$$V_{ext}^{elstat,n_{surr}} = \langle n_{surr}^{(0)} | V_{QM-surr} | n_{surr}^{(0)} \rangle, \quad (36)$$

is the external electrostatic potential created by the surroundings in the electronic state $n_{surr}$, $F_r^{QM,n_{QM}}(\mathbf{R}_I)$ is the electric field created in the surroundings at point $\mathbf{R}_I$ by the QM part in the quantum state $|n_{QM}^{(0)}\rangle$, $r_{rs}^{surr,I,n_{surr}}$ is the polarizability of the $I$-th CG site (molecule) in vacuum in the electronic state $n_{surr}$ defined in the usual way:

$$r_{rs}^{surr,I,n_{surr}} = \sum_{l_I \neq (n_{surr})_I} \frac{2\langle l_I^{(0)} | \tilde{\ }_{I,r}^{surr} | (n_{surr})_I^{(0)} \rangle \langle (n_{surr})_I^{(0)} | \tilde{\ }_{I,s}^{surr} | l_I^{(0)} \rangle}{E_{l_I}^{surr,I} - E_{(n_{surr})_I}^{surr,I}}, \quad (37)$$

and $s_{rs}^{surr,I,n_{surr}}$ is another characteristic of the $I$-th CG site (molecule) in vacuum in the electronic state $n_{surr}$ defined as follows:

$$s_{rs}^{surr,I,n_{surr}} = \sum_{l_I \neq (n_{surr})_I} \frac{2\langle l_I^{(0)} | \tilde{\ }_{I,r}^{surr} | (n_{surr})_I^{(0)} \rangle \langle (n_{surr})_I^{(0)} | \tilde{\ }_{I,s}^{surr} | l_I^{(0)} \rangle}{\left(E_{l_I}^{surr,I} - E_{(n_{surr})_I}^{surr,I}\right)^2}. \quad (38)$$

An outline of an algorithm for building a QM/CG-MM model with electrostatic and induction interactions between the QM part and the surroundings is presented in Scheme 2 of Appendix E later.

Finally, we note that in the case when only the electrostatic interaction and the polarization of the QM part by the unperturbed surroundings are taken into account, that is from the viewpoint analogous to the electrostatic embedding in the QM/MM theory,[27,28] significant simplifications are possible. This case is analyzed in Sec. 8 of the Supplemental Material. Briefly, the eigenvalues of the effective QM Hamiltonian assume the following form:

$$E_{eff,n_{QM}}^{(elstat+indQM)} = E_{n_{QM}}^{QM}(\mathbf{r}_n^{QM}) + V_{CG}^{surr}(\mathbf{R}^N) + \Delta E_{eff,n_{QM}}^{(elstat)}(\mathbf{r}_n^{QM}, \mathbf{R}^N) + \Delta E_{eff,n_{QM}}^{(indQM)}(\mathbf{r}_n^{QM}, \mathbf{R}^N), \quad (39)$$

where the induction contribution to the interaction energy $\Delta E_{eff,n_{QM}}^{(indQM)}$ is defined by:

$$\Delta E_{\text{eff},n_{QM}}^{(\text{ind QM})}\left(\mathbf{r}_{n}^{QM},\mathbf{R}^{N}\right)=-\frac{1}{S}\left[\ln\langle e^{-S\Delta E_{n_{QM},n_{surr}}^{(\text{elstat+ind QM})}(\mathbf{r}_{n}^{QM},\mathbf{r}_{n}^{surr})}\rangle_{\mathbf{R}^{N}}-\ln\langle e^{-S\Delta E_{n_{QM},n_{surr}}^{(\text{elstat})}(\mathbf{r}_{n}^{QM},\mathbf{r}_{n}^{surr})}\rangle_{\mathbf{R}^{N}}\right], \quad (40)$$

with

$$\Delta E_{n_{QM},n_{surr}}^{(\text{elstat+ind QM})}=\Delta E_{n_{QM},n_{surr}}^{(\text{elstat})}-\sum_{k\neq n_{QM}}\frac{\left|\langle k^{(0)}|V_{ext}^{elstat,n_{surr}}|n_{QM}^{(0)}\rangle\right|^{2}}{E_{k}^{QM}-E_{n_{QM}}^{QM}}+O(v^{3}). \quad (41)$$

The value of $\langle\hat{O}\rangle_{QM/CG-MM}$ can be computed in the following way:

$$\langle\hat{O}\rangle_{QM/CG-MM}=\frac{\int d\mathbf{r}_{n}^{QM}d\mathbf{R}^{N}\sum_{n_{QM}}e^{-SE_{\text{eff},n_{QM}}(\mathbf{r}_{n}^{QM},\mathbf{R}^{N})}\langle n_{QM}|\hat{O}|n_{QM}\rangle}{\int d\mathbf{r}_{n}^{QM}d\mathbf{R}^{N}\sum_{n_{QM}}e^{-SE_{\text{eff},n_{QM}}(\mathbf{r}_{n}^{QM},\mathbf{R}^{N})}}, \quad (42)$$

where the contributions to the eigenvalues of the effective Hamiltonian are restricted to the purely electrostatic term and the part of the induction due to the effect of the surroundings on the QM part, $E_{\text{eff},n_{QM}}=E_{\text{eff},n_{QM}}^{(\text{elstat+ind QM})}$, and the matrix elements of the operator $\hat{O}$ in the basis of the perturbed wavefunctions for the QM part are computed as follows:

$$\langle n_{QM}|\hat{O}|n_{QM}\rangle=\langle n_{QM}^{(0)}|\hat{O}|n_{QM}^{(0)}\rangle-\sum_{k\neq n_{QM}}c_{n_{QM},n_{surr}}^{k,n_{surr}}\left(\langle n_{QM}^{(0)}|\hat{O}|k^{(0)}\rangle+\langle k^{(0)}|\hat{O}|n_{QM}^{(0)}\rangle\right)+O(v^{2}). \quad (43)$$

A practical implementation of this approximation reduces to a simpler algorithm presented in Scheme 1 of Appendix E later.

**C. Dispersion interactions between the QM part and the surroundings**

We now add dispersion interactions between the QM part and the surroundings to electrostatic and induction interactions between the two parts. The resulting exact expressions for the coefficients $k_{i,j}^{n_{QM},n_{surr}}$ in terms of $c_{n_{QM},n_{surr}}^{k,l}$ are given in Sec. 5 of the Supplemental Material. To compute the missing values of $c_{n_{QM},n_{surr}}^{k,l}$, one needs to solve a problem similar to that discussed in Sec. V.B, namely to calculate $\Delta E_{n_{QM},n_{surr}}^{(disp)}$ and $k_{i,j}^{(disp)n_{QM},n_{surr}}$ without using wave functions of the surroundings. A possible approach to solving this problem is suggested in Sec. 9 of the Supplemental Material. Briefly, the dispersion contribution to the fine-grained interaction energy can be found as

$$\Delta E_{n_{QM},n_{surr}}^{(disp)} = -\frac{1}{2} \sum_{k \neq n_{QM}} \sum_{I \in CG} \sum_{\substack{r=x,y,z \\ s=x,y,z}} F_r^{QM,k,n_{QM}}(\mathbf{R}_I) \Gamma_{rs}^{surr,I,n_{surr}} F_s^{QM,k,n_{QM}}(\mathbf{R}_I) + O(v^3), \quad (44)$$

while the sums $\sum_{l \neq n_{surr}} c_{n_{QM},n_{surr}}^{i,l} c_{n_{QM},n_{surr}}^{j,l}$ required to compute the coefficients $k_{i,j}^{(ind+disp)n_{QM},n_{surr}}$ can be found in the following way:

$$\sum_{l \neq n_{surr}} c_{n_{QM},n_{surr}}^{i,l} c_{n_{QM},n_{surr}}^{j,l} = \frac{1}{2} \sum_{I \in CG} \sum_{\substack{r=x,y,z \\ s=x,y,z}} F_r^{QM,i,n_{QM}}(\mathbf{R}_I) S_{rs}^{surr,I,n_{surr}} F_s^{QM,j,n_{QM}}(\mathbf{R}_I) + O(v^3). \quad (45)$$

Here

$$F_r^{QM,k,n_{QM}}(\mathbf{R}_I) = \langle k^{(0)} | F_r^{QM}(\mathbf{R}_I) | n_{QM}^{(0)} \rangle \quad (46)$$

is an off-diagonal generalization of the above-mentioned variable $F_r^{QM,n_{QM}}(\mathbf{R}_I)$.

The algorithm presented in Scheme 2 of Appendix E implements the version of the QM/CG-MM approach accounting for the electrostatic, induction, and dispersion interactions between the QM part and the surroundings.

**D. Purely exchange interactions between the QM part and the surroundings**

In Secs. V.A-V.C, all types of exchange interactions between the two parts of the system were neglected. This subsection addresses purely exchange interactions, that is the interactions that yield the contributions to the effective QM Hamiltonian coming from the indistinguishability of all electrons in the system, in the absence of electrostatic, induction, and dispersion interactions. The contributions stemming from a combination of exchange interaction with electrostatic, induction, and dispersion interactions are discussed below in Sec. V.E.

As is known from the perturbative theory of intermolecular interactions, the purely exchange contribution to the interaction energy vanishes:

$$\Delta E_{n_{QM},n_{surr}}^{(pur.exch)} = \lim_{v \to 0} \left( \frac{\langle n_{QM}^{(0)} n_{surr}^{(0)} | \mathcal{A}_{sys} \hat{H}_e \mathcal{A}_{sys} | n_{QM}^{(0)} n_{surr}^{(0)} \rangle}{\langle n_{QM}^{(0)} n_{surr}^{(0)} | \mathcal{A}_{sys} \mathcal{A}_{sys} | n_{QM}^{(0)} n_{surr}^{(0)} \rangle} - E_{n_{QM}}^{QM} - E_{n_{surr}}^{surr} \right)$$
$$= \lim_{v \to 0} \frac{\langle n_{QM}^{(0)} n_{surr}^{(0)} | \mathcal{A}_{sys} V_{QM-surr} | n_{QM}^{(0)} n_{surr}^{(0)} \rangle}{\langle n_{QM}^{(0)} n_{surr}^{(0)} | \mathcal{A}_{sys} | n_{QM}^{(0)} n_{surr}^{(0)} \rangle} = 0. \quad (47)$$

However, purely exchange contributions to the coefficients $k_{i,j}^{n_{QM},n_{surr}}$, and thus to the eigenvectors and eigenvalues of the effective QM Hamiltonian $E_{eff,n_{QM}}$, are nontrivial. The corresponding analysis is technically involved and is found in the Supplemental Material, Sec. 10.

Interestingly, the purely exchange interaction affects the effective QM Hamiltonian $\hat{H}_{eff}^{QM}$ in QM/CG-MM models in a way different from the effect of exchange interactions in perturbative theories of intermolecular interactions. To demonstrate this, we consider the lowest order nontrivial correction to the lowest eigenvalue of this Hamiltonian $E_{eff,0}$. As follows from Eq. (19) of the main text and Eq. (S61) of the Supplemental Material, in the perturbative regime,

$$\Delta E_{eff,0}^{(pur.exch.)}\left(\mathbf{r}_n^{QM}, \mathbf{R}^N\right) = \frac{\ln C_{N_e^{sys}}^{N_e^{QM}}}{s}$$
$$+ \frac{N_e^{QM} N_e^{surr}}{s} \langle \int d\mathbf{x} d\mathbf{x}' \ldots_{n_{surr}}^{surr}(\mathbf{x},\mathbf{x}') \ldots_{0}^{QM}(\mathbf{x},\mathbf{x}') \rangle_{\mathbf{R}^N} + O(s^3). \quad (48)$$

where $s$ is a small parameter characterizing a degree of overlap of molecular orbitals in the QM part and the surroundings (strictly defined in Sec. 10 of the Supplemental Material), $C_{N_e^{sys}}^{N_e^{QM}} = N_e^{sys}!/(N_e^{QM}!N_e^{surr}!)$ is a binomial coefficient, $\ldots_{n_{surr}}^{surr}$ and $\ldots_{n_{QM}}^{QM}$ are one-particle off-diagonal electron density matrices for the surroundings and the QM part, respectively, and **x, x'** are two three-dimensional vectors representing coordinates of an electron not integrated out in the one-particle densities $\ldots_{n_{surr}}^{surr}$ and $\ldots_{n_{QM}}^{QM}$. The first term on the right hand side of Eq. (48) is the same for all eigenvalues of the effective QM Hamiltonian, and its only effect is to shift the energy scale of $E_{eff, n_{QM}}$ by the value of $k_B T \ln C_{N_e^{sys}}^{N_e^{QM}}$. The second term, which has the order of $O(s^2)$, is specific to $E_{eff,0}$; corrections to the other eigenvalues of the effective Hamiltonian have a different functional form. The key observation is that this second term is proportional to temperature of the system, unlike various sorts of exchange energies (e.g., the exchange repulsion energy) in the theory of intermolecular interactions. The temperature dependence of this term comes from the thermodynamic averaging implied by coarse-graining and does not have an analogue in the theory of intermolecular interactions.

**E. Exchange-repulsion, exchange-induction, and exchange-dispersion interactions between the QM part and the surroundings**

As can be seen from Eq. (19), combinations of exchange interaction with electrostatic, induction, and dispersion interactions affect the effective QM Hamiltonian via two different ways: (1) via the fine-grained energy of interaction between the QM part and the surroundings $\Delta E_{n_{QM},n_{surr}}$, and (2) via the coefficients $k_{i,j}^{n_{QM},n_{surr}}$. The resulting contributions to the eigenvalues and eigenfunctions of the effective QM Hamiltonian can be defined, as usual, by a subtractive scheme. For example, the exchange-repulsion contribution to the $n_{QM}$-th eigenvalue is

$$\Delta E_{eff,n_{QM}}^{(exch.-rep.)}\left(\mathbf{r}_n^{QM}, \mathbf{R}^N\right) = E_{eff,n_{QM}}^{(exch.-rep.)}\left(\mathbf{r}_n^{QM}, \mathbf{R}^N\right) - E_{eff,n_{QM}}^{(pur.exch.)}\left(\mathbf{r}_n^{QM}, \mathbf{R}^N\right), \tag{49}$$

where $E_{eff,n_{QM}}^{(\text{exch.-rep.})}$ is computed by diagonalization of the matrix given by Eq. (19) using the values of $k_{i,j}^{n_{QM},n_{surr}} = k_{i,j}^{(\text{pur.exch.})n_{QM},n_{surr}}$ given by Eq. (S58) of the Supplemental Material [since the electrostatic interaction does not explicitly affects $k_{i,j}^{n_{QM},n_{surr}}$] and $\Delta E_{n_{QM},n_{surr}} = \Delta E_{n_{QM},n_{surr}}^{(\text{elstat})}$ given by Eq. (59) [since the purely exchange interaction does not contribute to $\Delta E_{n_{QM},n_{surr}}$, Eq. (47)], and $E_{eff,n_{QM}}^{(\text{pur.exch.})}$ is computed by diagonalization of the matrix given by Eq. (19) using the values of $k_{i,j}^{n_{QM},n_{surr}} = k_{i,j}^{(\text{pur.exch.})n_{QM},n_{surr}}$ and $\Delta E_{n_{QM},n_{surr}} = 0$.

Explicit analytical expressions for $\Delta E_{eff,n_{QM}}^{(\text{exch.-rep.})}$, $\Delta E_{eff,n_{QM}}^{(\text{exch.-ind.})}$, and $\Delta E_{eff,n_{QM}}^{(\text{exch.-disp.})}$ would be too sophisticated (not given), and numerical analysis of these contributions based on decompositions similar to that given by Eq. (49) may prove to be more efficient in practice.

**F. Overview of the contributions to the effective Hamiltonian**

*i. Decomposition of the contributions*

The analysis performed in this Sec. V represents each eigenvalue $E_{eff,n_{QM}}$ of the effective QM Hamiltonian $\hat{H}_{eff}^{QM}$ as the sum of the corresponding energy level of the isolated QM part $E_{n_{QM}}^{QM}$, the free energy of the isolated surroundings $V_{CG}^{surr}$, and contributions stemming from different types of interactions between the QM part and the surroundings, namely

$$\begin{aligned}
E_{eff,n_{QM}}\left(\mathbf{r}_n^{QM}, \mathbf{R}^N\right) &= E_{n_{QM}}^{QM}\left(\mathbf{r}_n^{QM}\right) + V_{CG}^{surr}\left(\mathbf{R}^N\right) + \Delta E_{eff,n_{QM}}^{(\text{elstat})}\left(\mathbf{r}_n^{QM}, \mathbf{R}^N\right) \\
&+ \Delta E_{eff,n_{QM}}^{(\text{indQM})}\left(\mathbf{r}_n^{QM}, \mathbf{R}^N\right) + \Delta E_{eff,n_{QM}}^{(\text{indCG})}\left(\mathbf{r}_n^{QM}, \mathbf{R}^N\right) + \Delta E_{eff,n_{QM}}^{(\text{disp})}\left(\mathbf{r}_n^{QM}, \mathbf{R}^N\right) \\
&+ \Delta E_{eff,n_{QM}}^{(\text{pur.exch.})}\left(\mathbf{r}_n^{QM}, \mathbf{R}^N\right) + \Delta E_{eff,n_{QM}}^{(\text{exch.-rep.})}\left(\mathbf{r}_n^{QM}, \mathbf{R}^N\right) \\
&+ \Delta E_{eff,n_{QM}}^{(\text{exch.-ind.})}\left(\mathbf{r}_n^{QM}, \mathbf{R}^N\right) + \Delta E_{eff,n_{QM}}^{(\text{exch.-disp.})}\left(\mathbf{r}_n^{QM}, \mathbf{R}^N\right).
\end{aligned} \quad (50)$$

Explicit expressions for $\Delta E_{eff,n_{QM}}^{(elstat)}$, $\Delta E_{eff,n_{QM}}^{(indQM)}$ and $\Delta E_{eff,n_{QM}}^{(exch.-rep.)}$ are given by Eqs. (25), (40) and (49), while the other contributions can be written in a way analogous to Eq. (49).

Similar decomposition can be written for the eigenfunctions of the effective QM Hamiltonian. Using these decompositions for the eigenvalues and the eigenfunctions, the partition function $Z$ for the full system can be computed by Eq. (28), and expectation values of QM operators $\langle \hat{O} \rangle_{QM/CG-MM}$ by Eq. (42).

Explicit expressions for the contributions of several first types of interactions [e.g., Eq. (25) of the main text and Eq. (S26) of the Supplemental Material for $\Delta E_{eff,n_{QM}}^{(elstat)}$, or Eq. (40) of the main text and Eq. (S49) of the Supplemental Material for the part of $\Delta E_{eff,n_{QM}}^{(ind)}$ coming from the polarization of the QM part by the surroundings] are relatively easy and informative. On the other hand, explicit expressions for later contributions are not so simple, if available in an analytical form at all.

*ii. Limitations of the perturbative analysis*

The main difficulty in a perturbative estimation of the latter contributions on the right hand side of Eq. (50) is caused by the fact that for the excited states of the QM part, the second term on the right hand side of Eq. (19) may not serve as a small correction to the first term even if the coefficients $k_{i,j}^{n_{QM},n_{surr}}$ are small. Indeed, the ratio of the second term to the corresponding diagonal value of the first term equals

$$\sum_{n_{QM}} \frac{e^{-\mathrm{s} E_{n_{QM}}^{QM}(\mathbf{r}_n^{QM})} \langle k_{i,j}^{n_{QM},n_{surr}}(\mathbf{r}_n^{QM},\mathbf{r}_n^{surr}) e^{-\mathrm{s}\Delta E_{n_{QM},n_{surr}}(\mathbf{r}_n^{QM},\mathbf{r}_n^{surr})} \rangle_{\mathbf{R}^N}}{e^{-\mathrm{s} E_i^{QM}(\mathbf{r}_n^{QM})} \langle e^{-\mathrm{s}\Delta E_{n_{QM},n_{surr}}(\mathbf{r}_n^{QM},\mathbf{r}_n^{surr})} \rangle_{\mathbf{R}^N}}, \qquad (51)$$

which is the sum of terms on the order of $O\left(e^{-S\left(E_{n_{QM}}^{QM}(\mathbf{r}_n^{QM})-E_i^{QM}(\mathbf{r}_n^{QM})\right)}\right)O\left(k_{i,j}^{n_{QM},n_{surr}}\right)$. In the case of excited states, the terms for $n_{QM}<i$ may prove to be large even if $k_{i,j}^{n_{QM},n_{surr}}\ll 1$ because of the factor $e^{-S\left(E_{n_{QM}}^{QM}(\mathbf{r}_n^{QM})-E_i^{QM}(\mathbf{r}_n^{QM})\right)}>1$. This fact does not undermine the usefulness of the perturbative expansions for the values of $\Delta E_{n_{QM},n_{surr}}$ and $k_{i,j}^{n_{QM},n_{surr}}$ discussed above in this section, because these expansions still can be used to compute the matrix elements $\left\langle i^{(0)}\left|e^{-S\hat{H}_{eff}^{QM}}\right|j^{(0)}\right\rangle$ in a perturbative regime. After that, a numerical diagonalization of this matrix will yield the eigenvalues and eigenfunctions of the effective QM Hamiltonian. It is only the stage of the matrix diagonalization that cannot be done analytically in a perturbative manner.

*iii. Additional remarks*

The presented framework reaches the goal of building the "average-then-interact" picture of a heterogeneous molecular system. In simplest cases, "averaging" (that is, CG-ing of the surroundings) is possible in an analytical form, as in Eqs. (S26) or (S49) of the Supplemental Material, directly yielding an analytical dependence of the effective QM Hamiltonian on the CG variables $\mathbf{R}^N$. In a more general scenario, such averaging is possible in a numerical form, based on molecular dynamics simulation of the surroundings. After this averaging has been performed, the "interaction" (between the QM part and the surroundings) can be analyzed in a more efficient and conceptually simpler way.

Depending on the nature of the system under investigation and the demanded accuracy level, some of the terms in Eqs. (50) and its analogue for the eigenfunctions may be omitted, leading to a variety of possible QM/CG-MM models.

## VI. DISCUSSION

### A. Overview of the QM/CG-MM approach

The current paper presents a strict *ab initio* "bottom-up" derivation of the QM/CG-MM approach. This derivation is based on the exact quantum description of the entire system and on the use of a consistency condition ensuring that a QM/CG-MM model makes, in principle (that is, in the limit of infinite basis set and complete statistical sampling), exact predictions about the observable properties of a system under investigation at equilibrium.

Though at first the consistency condition is introduced in this paper only for the quantum mechanical observables for the QM part, later it is demonstrated that exact predictions of the classical observables for the surroundings on the CG resolution level are also ensured by the QM/CG-MM model. Two basic results presented in this paper are Eq. (8), which determines the effective Hamiltonian for the QM part as affected by the surroundings, and Eq. (15), which, conversely, describes the effect of the QM part on the surroundings. As follows from these two equations, the central problem to be solved in the theory of QM/CG-MM modeling is to compute the effective Hamiltonian for the QM part.

In some cases, CG of the surroundings can be performed analytically, leading to explicit analytical expressions for the effective Hamiltonian of the QM part in terms of CG variables, such as those following from Eqs. (S26) or (S49) of the Supplemental Material. For real-world complex systems of practical interest and when high accuracy of a model is demanded, a numerical computation of the effective Hamiltonian, with explicit MM sampling of the surroundings at atomistic level, is always an available option.

Different versions of QM/CG-MM models can exist, based on neglecting some of the terms on the right hand side of Eq. (50) and on the use of different approximations for the remaining terms. Some of these versions are presented in Schemes 1 and 2 in Appendix E of this paper. For various molecular systems to be studied and various required accuracy levels, different versions of QM/CG-MM may appear optimal.

**B. Relationship of QM/CG-MM to QM/MM**

The QM/CG-MM and QM/MM approaches share a number of common features. First, from the viewpoint of interaction energies QM/CG-MM models are similar to QM/MM ones with additive coupling schemes. For example, $\Delta E^{(elstat)}_{eff,n_{QM}}$ given by Eq. (25) is analogous to the QM-MM interaction potential $V_{QM-MM}$ in QM/MM models with mechanical embedding. At the same time, the method of computing the averages $\langle \cdots \rangle_{\mathbf{R}^N}$ suggested in Appendix C resembles subtractive coupling in QM/MM models, where, unlike in models with additive coupling, MM force field for the QM part needs to be defined. However, when the dispersion and exchange interactions in a QM/CG-MM model are accounted for, analytical expressions for some of the $\Delta E$ terms in Eq. (50) are not available, building the effective Hamiltonian requires a numerical diagonalization of the matrix computed by Eq. (19), and the analogy between such advanced QM/CG-MM schemes and additive or subtractive QM/MM coupling schemes vanishes.

Second, the problem of covalent bonds crossing the border between the QM and MM part in QM/MM models is entirely inherited by QM/CG-MM models. In general, one may suggest using in QM/CG-MM models the same techniques as those used in the QM/MM models (link atoms, localized orbitals, etc.). These techniques seem to be compatible only with a numerical "average-then-interact" approach, since they require an intermediate atomistic representation of the surroundings to couple the QM and CG parts. A recent strict analysis of the problem of the QM/MM interface[29] using the toolkit of density matrix embedding theory (DMET) may probably be extended to dealing with covalent bonds crossing the QM/CG-MM interface.

An important, deep difference between the QM/CG-MM and QM/MM paradigms is that in the former approach the temperature of the surroundings is explicitly defined. Therefore, the QM/CG-MM description is more straightforwardly related to the experimental settings in most practically important heterogeneous molecular systems. Other advantages of the QM/CG-MM methodology over the QM/MM one include lower computational cost, simplicity of interpretation, and more complete sampling of the configuration space, all of these being inherited from the CG methodology.

C. Relationship of QM/CG-MM to other hybrid methods

Density matrix embedding theory (DMET)[30,31] is another exact embedding scheme, different from the QM/MM approach. The main difference of the proposed QM/CG-MM methodology from DMET is that the former one deals with thermodynamic distributions; in particular, a QM/CG-MM model implies a well-defined temperature of the environment.

The effective fragment potential (EFP) method,[32-34] similarly to the QM/CG-MM approach, includes the analysis of different (electrostatic, induction, dispersion, exchange) contributions to the interfragment interaction energy; introduces effective repulsion potentials to account for Pauli repulsion; implies using preliminary QM computations for molecules in the surroundings to estimate the charge distribution and polarizability. Two most important differences between the QM/CG-MM and EFP methodologies are that in QM/CG-MM models a coarser representation of the surroundings is used, and the temperature of the surroundings is explicitly defined.

In the QM/(AA+CG) approach,[23] the environment is modeled with an adaptive AA+CG scheme that combines all-atom and CG representations. The energy of interaction between the QM and (AA+CG) parts is written by analogy with the potential used in QM/MM models with electrostatic embedding. In doing so, the terms in a QM/MM potential corresponding to all-atom representation are replaced by analogous terms written in a mixed (AA+CG) representation. No derivation of the potential from the first principles is provided, leaving the question of limitations of this approach open, which differs it from the systematic bottom-up QM/CG-MM method.

The problem of a strict derivation of the effective potential of interactions between subsystems refers not only to the interactions between the QM part and classical surroundings. In various AA/CG schemes, interactions between the atomically resolved and CG parts of the system are also introduced *ad hoc*.[35] By contrast, the QM/CG-MM inherits a strict bottom-up approach to the coarse-graining of the classical part of a modeled system from the MS-CG methodology.

The coarse-grained QM/MM model presented in ref. [20] is closer to continuous solvent models,[36,37] since different parts of the surroundings are defined by a dissection of the space into constant regions, while the approach presented in this article takes in account the chemical nature of the surroundings (in particular, can treat a molecule in the surroundings as the same CG site at different moments in time, despite that the molecule moves in space). The averaged condensed phase environment (ACPE) model later developed in the same group[24] accounts for specific locations and nature of atoms and molecules in the surroundings when coarse-graining is performed. However, the procedure of configurational averaging used to build the effective Hamiltonian is still introduced without a strict derivation from first principles.

To summarize, the QM/CG-MM methodology proposed in this paper, despite some overlap with the ideas from the QM/MM, QM/(AA+CG), EFP, DMET and other methods, [19] is an original systematic approach to modeling complex molecular systems.

### D. Fundamental limitations to the QM/CG-MM method

First of all, we note that at the very beginning of Sec. II the Born-Oppenheimer approximation was introduced for both the QM part and the surroundings. The reason was that the use of the technique of CG required a classical description of the atoms[16] (in this context, nuclei) in the surroundings. Recently, the CG methodology has been generalized to quantum systems.[38] In the present work though, we focus on modeling systems without non-adiabatic transitions in the surroundings. As for the restriction to non-adiabatic processes in the QM part, it seems less obligatory from the conceptual viewpoint, but useful for notational convenience. We leave the problem of QM/CG-MM modeling of systems with strong non-adiabatic effects in the QM region for future work.

Neglect of charge transfer between the QM part and the surroundings is another approximation used in Sec. II. It seems possible to extend the QM/CG-MM methodology to the case of non-negligible charge transfer by using the formalism of second quantization, as outlined in Appendix A. We also leave this generalization for future work. For the present, we notice that in many practically important cases the issue of charge transfer may be circumvented by redefining the boundary between the parts of the system such that all donor and acceptor atoms or molecules are localized to the QM region.

The question of the applicability of the present approach to study the dynamics of molecular systems also remains open. Classical CG models are known to overestimate the rate of conformational changes unless special measures (e.g., introduction of friction forces) are undertaken.[39,40] Practically efficient approaches to modeling quantum dynamics of molecular systems usually rely on the linear response theory that relates time correlation functions to expectation values of certain operators at thermodynamic equilibrium.[41,42] Possible ways of combining these or any other alternative approaches to the description of the dynamics of molecular systems in the QM/CG-MM methodology go beyond the scope of this work.

## VII. CONCLUSIONS

The present article, to the best of our knowledge, formalizes for the first time the idea of QM/CG-MM modeling and demonstrates that this approach is in principle exact (in the sense defined in Sec. II, with additional clarifications given in Appendix B).

The proposed method occupies a niche between QM/MM and QM/continuum approaches. It is valuable when a representation of the environment at the MM level does not provide sufficient sampling of the corresponding thermodynamic ensemble and/or cannot be simply interpreted due to the vastness of possible states of the MM part, while a representation of the environment as a continuous medium with a well-defined dielectric permittivity is too coarse. We expect that the QM/CG-MM methodology will prove to be optimal for modeling multiple practically important complex molecular systems, including enzymes and composite materials.

This paper introduces a hierarchy of approximations that can be used in practice to compute the effective Hamiltonian, leading to various versions of QM/CG-MM models with different accuracy level and computational cost. In particular, accounting only for the electrostatic interaction between the QM part and the surroundings, as described in Sec. V.A, yields a simplest QM/CG-MM model that can be practically implemented based on the algorithm given in Scheme 1 of Appendix E. Accounting for induction and dispersion interactions, as described in Sec. V.B-V.C, leads to more advanced QM/CG-MM models, which require a more sophisticated algorithm such as one given in Scheme 2 of Appendix E. Various exchange interactions can also be accounted for in QM/CG-MM models, as discussed in Sec. V.D-V.E.

Future directions of this research may include the following: Some improvements of the methodology could be possible, providing for partial easing of the general assumptions used in this article (no charge transfer, Born-Oppenheimer approximation for the QM part, perturbative treatment of excited states) or the use of the QM/CG-MM model as a part of bigger mixed-resolution models, e.g. QM/MM/CG[22] or QM/CG/continuum models. A combination of the QM/MM and ultra-coarse-graining (UCG)[43-45] ideas may also be fruitful, leading to a QM/UCG approach allowing for different discrete states of the surroundings. Application of the presented QM/CG-MM methodology to various molecular systems and comparison of the predictions of the QM/CG-MM models with experimental data or other computational methods will be required in the future. Specific improvements in the QM/CG-MM machinery, for example further development of the techniques for treatment of exchange interactions or dealing with covalent bonds crossing the boundary between the QM and CG regions, are also desirable. New approaches (including non-perturbative ones) to computing the eigenfunctions and eigenvalues of the effective Hamiltonian may also prove fruitful.

To conclude, this paper opens up a new perspective in computational chemistry by combining the well-known ideas of QM/MM modeling and coarse-graining into a single, synergistic, and "bottom-up" QM/CG-MM approach.

**SUPPLEMENTARY MATERIAL**

See Supplemental Material for technical details on derivations of some of the results presented in the main text and appendices.

**ACKNOWLEDGMENTS**

This research was supported by the Office of Naval Research (ONR Grant No. N00014-15-1-2493) in part by the National Science Foundation (NSF Grant CHE-1465248). The authors are grateful to Jason Elangbam, Dr. Olaseni Sode, Dr. James F. Dama, and Dr. Daniel Silverstein for valuable discussions at various stages of the project and comments on the manuscript.

**APPENDICES**

**A. Charge transfer between the QM part and the surroundings in QM/CG-MM models.**

In general, the number of electrons in the QM part $N_e^{QM}$ is not a conserved quantum number, as long as the QM part and the surroundings interact with each other. A proper way to describe possible electronic states of the QM part in this situation is in terms of states in the Fock space $\mathscr{F}$, defined as the direct sum of antisymmetric tensor powers of a single-particle Hilbert space $\mathscr{H}$,[46]

$$\mathscr{F} = \bigoplus_{N_e^{QM}=0}^{N_e^{sys}} \mathcal{A}\mathscr{H}^{\otimes N_e^{QM}}, \tag{52}$$

with the number of electrons (the powers of the single-particle Hilbert space) $N_e^{QM}$ running from 0 to $N_e^{sys}$. Respectively, the effective Hamiltonian for the QM part can be written employing the formalism of second quantization, with the amplitudes depending on the coordinates of the nuclei in the QM part and the CG variables characterizing the surroundings.

To illustrate this idea, consider a molecular system AB with charge transfer, following the classical work by Mulliken.[47] He states that if A has a high electron affinity and B has a low ionization potential, then the wave function of the system can be written as

$$\Psi_{sys} = a\mathcal{A}\big(\Psi(A)\Psi(B)\big) + b\mathcal{A}\big(\Psi(A^-)\Psi(B^+)\big) + \cdots, \tag{53}$$

where $a$ and $b$ are coefficients, $\mathcal{A}$ is the antisymmetrization operator and $\cdots$ represents "small modifying terms".[47] Now, let us interpret A as the QM part and B as the surroundings. (Interestingly, though A and B are neutral or charged atoms or molecules in all of the specific cases discussed by Mulliken, he mentions that A and B may be "perhaps even solids", media capable of achieving thermodynamic equilibrium.) In accordance to the discussion above, the full wave function of the system $\Psi_{sys}$ belongs to a $N_e^{sys}$–particle Hilbert space, where $N_e^{sys}$ is the total number of the electrons in AB. However, the description of only the QM part, abstracting from the surroundings, requires the use of the set of wave functions $\Psi(A)$ and $\Psi(A^-)$ [and possibly some other wave functions involved in the "small modifying terms" in Eq. (53)] that correspond to different numbers of electrons in the QM part and belong to the Fock space $\mathscr{F}$ defined by Eq. (52).

Hence, a strict formulation of the QM/CG-MM methodology for the case of an undefined number of electrons in the QM part (in other words, non-negligible charge transfer between the QM part and the surroundings) appears to be possible in terms of the Fock space and second quantization. We leave a detailed elaboration in this direction for future investigation. In this paper, we note that, in most practically important cases, it may make sense to choose the boundary between the QM and CG parts of the system in such a way that charge transfer across this border is negligible. The class of cases is analyzed in the main text.

**B. Localization of observable operators to the QM part.**

It is instructive to start this discussion with one important methodological observation about the QM/MM approach. QM/MM models are typically used to obtain information about the QM part of a system under study. Technically, this is done by computing expectation values of certain quantum mechanical operators. Note that an arbitrary quantum mechanical operator of interest $\hat{O}$ in a QM/MM model acts only on the electronic degrees of freedom associated with the QM part, but not the MM part: $\hat{O} = \hat{O}(\mathbf{r}_e^{QM})$. On the other hand, operators for physical observables must commute with the antisymmetrization operator $\mathscr{A}_{sys}$ for all the electrons in the system (that is, both in the QM and MM parts), otherwise physical experiments measuring this observable would be able to distinguish some electrons from others. Evidently, this condition cannot be satisfied for an arbitrary operator $\hat{O}(\mathbf{r}_e^{QM})$ that does not act on the coordinates of electrons in the MM part, and therefore the QM/MM models yield an expectation values of operators that are, strictly speaking, not physical. However, in the approximation of weak exchange interactions between the QM and MM parts, expectation values of operators $\hat{O}(\mathbf{r}_e^{QM})$ localized to the QM part coincide with expectation values of the physical operators obtained by antisymmetrization of $\hat{O}(\mathbf{r}_e^{QM})$, namely $\mathcal{A}_{sys}\hat{O}(\mathbf{r}_e^{QM})\mathcal{A}_{sys}$.

This problem of localization of quantum mechanical operators actually goes beyond the QM/MM approach. In purely quantum computations, models of molecular systems under investigation are often built is such a way that they contain a finite, relatively small, number of electrons (e.g., a molecule in vacuum). However, results of the computations (specifically, expectation values of localized quantum mechanical operators) are then compared to experimental results that refer to a system of interest embedded into much larger environment ("the rest of the universe"). If exchange interactions between the system explicitly included into the model and the environment is weak (e.g., as in the case of molecules in a low-density gas), then this replacement of a physical operator by an operator localized to the system explicitly described in QM terms is a reasonable approximation.

Returning to the QM/CG-MM approach, we claim that, strictly speaking, it is impossible to find a definition of the effective QM Hamiltonian for a system of indistinguishable particles that would satisfy Eq. (7) in the general case of an arbitrary operator $\hat{O}$, for the following reason: As in the case of the QM/MM approach, the computation of an expectation value of an arbitrary operator $\hat{O}$ from the QM/CG-MM model makes sense only if $\hat{O}$ acts only on the electronic degrees of freedom associated with the QM part, though it may also parametrically depend on the coordinates of the nuclei in the QM part:

$$\hat{O} = \hat{O}\left(\mathbf{r}_e^{QM}; \mathbf{r}_n^{QM}\right). \tag{54}$$

On the other hand, operators for physical observables must commute with the antisymmetrization operator for *all* the electrons in the system:

$$\left[\hat{O}, \mathcal{A}_{sys}\right] = 0. \tag{55}$$

Evidently, Eq. (54) contradicts Eq. (55). This contradiction leads to the impossibility of satisfying condition (7) using the effective QM Hamiltonian given by Eq. (8).

However, we claim that the definition of the effective QM Hamiltonian by Eq. (8), with the interpretation of the trace given by Eq. (10) and the value of $C$ given by Eq. (12), is the optimal one and should be used in the QM/CG-MM approach. To support this claim, consider the following expression for the difference between $\langle\hat{O}\rangle_{QM/CG-MM}$ and $\langle\hat{O}\rangle$ that can be derived from Eqs. (2), (5), (8), and (10) – (12):

$$\langle\hat{O}\rangle_{QM/CG-MM} - \langle\hat{O}\rangle = -\frac{1}{Z}\int\left\{\frac{d\mathbf{r}_n d\mathbf{p}_n}{(2f\hbar)^{3N_n^{sys}}} e^{-\beta T_n(\mathbf{p}_n)}\right. \tag{56}$$
$$\left.\times\left(N_e^{sys}!\right)\int d\mathbf{r}_e^{QM} d\mathbf{r}_e^{surr} \left\langle \mathbf{r}_e^{QM}, \mathbf{r}_e^{surr} \left| \mathcal{A}_{sys} e^{-\beta \hat{H}_e}[\hat{O}, \mathcal{A}_{sys}]\mathcal{A}_{QM} \right| \mathbf{r}_e^{QM}, \mathbf{r}_e^{surr}\right\rangle\right\}.$$

In this equation, the only source of the difference between $\langle\hat{O}\rangle_{QM/CG-MM}$ and $\langle\hat{O}\rangle$ appears to come from the commutator of the operator $\hat{O}\left(\mathbf{r}_e^{QM}; \mathbf{r}_n^{QM}\right)$ with the full antisymmetrizer $\mathcal{A}_{sys}$.

Thus, the indistinguishability of electrons in the QM part and the surroundings makes observable physical quantities characterizing the QM part somewhat uncertain. However, this should not be interpreted as a defect of the QM/CG-MM methodology, because the same problem arises in any purely QM or QM/MM model: a molecular system in vacuum or an external field or, respectively, an active part of a system, do in principle exchange electrons with the rest of the universe. The uncertainty in the characteristics of the QM part can be minimized by a proper choice of the boundary between the QM part and the surroundings.

**C. On efficient computation of averages defined by Eq. (21).**

Averaging $\langle \cdots \rangle_{\mathbf{R}^N}$ occurs in a key equation, Eq. (19), as well as numerous subsequent equations derived from Eq. (19). Direct computation of averages $\langle \cdots \rangle_{\mathbf{R}^N}$ based on their definition, Eq. (21), is computationally impractical. To see this, note that the probability density to find a specific configuration of the molecules in the surroundings, according to Eq. (21), is the same as in the surroundings in the absence of the QM part (e.g., in a pure solvent without a solute). However, typical configurations of the surroundings will correspond to extremely high positive values of $\Delta E_{n_{QM}, n_{surr}}$ due to overlap of the molecules from the QM part and the surroundings, and the corresponding contributions to $\langle e^{-S \Delta E_{n_{QM}, n_{surr}}} \rangle_{\mathbf{R}^N}$ and $\langle k_{i,j}^{n_{QM}, n_{surr}} e^{-S \Delta E_{n_{QM}, n_{surr}}} \rangle_{\mathbf{R}^N}$ will be unreasonably small. On the other hand, moderate values of $\Delta E_{n_{QM}, n_{surr}}$ correspond to the configurations of the surroundings with a cavity that would fit the QM part in, but the probability to find such a configuration in the ensemble used for the averaging $\langle \cdots \rangle_{\mathbf{R}^N}$ is extremely small since it equals the probability of spontaneous formation of a relatively large cavity in the isolated surroundings (e.g., in a pure solvent). A simple practical way to compute averages $\langle \cdots \rangle_{\mathbf{R}^N}$ may be to use the following identity following from Eq. (21):

$$\langle g(\mathbf{r}_n^{QM}, \mathbf{r}_n^{surr}) \cdot e^{-S \Delta E_{n_{QM}, n_{surr}}(\mathbf{r}_n^{QM}, \mathbf{r}_n^{surr})} \rangle_{\mathbf{R}^N} = \frac{\langle g(\mathbf{r}_n^{QM}, \mathbf{r}_n^{surr}) \cdot e^{-S \left[ \Delta E_{n_{QM}, n_{surr}}(\mathbf{r}_n^{QM}, \mathbf{r}_n^{surr}) - \mathcal{E}(\mathbf{r}_n^{QM}, \mathbf{r}_n^{surr}) \right]} \rangle_{\mathbf{R}^N, ac}}{\langle e^{S \mathcal{E}(\mathbf{r}_n^{QM}, \mathbf{r}_n^{surr})} \rangle_{\mathbf{R}^N, ac}}, \quad (57)$$

where $\mathscr{E}\left(\mathbf{r}_n^{QM},\mathbf{r}_n^{surr}\right)$ is a computationally inexpensive (e.g., classical force field) approximation for the interaction energy and $g\left(\mathbf{r}_n^{QM},\mathbf{r}_n^{surr}\right)$ is an arbitrary function of the nuclear coordinates. In Eq. (57), $\langle\cdots\rangle_{\mathbf{R}^N,ac}$ stands for averaging defined in the following way:

$$\langle f\rangle_{\mathbf{R}^N,ac} = \frac{\sum_{n_{surr}}\int d\mathbf{r}_n^{surr}\, \sqcup\left(\mathbf{R}^N - M_R^N(\mathbf{r}_n^{surr})\right)e^{-\mathrm{s}\left[E_{n_{surr}}^{surr}(\mathbf{r}_n^{surr})+\mathscr{E}(\mathbf{r}_n^{QM},\mathbf{r}_n^{surr})\right]}f}{\sum_{n_{surr}}\int d\mathbf{r}_n^{surr}\, \sqcup\left(\mathbf{R}^N - M_R^N(\mathbf{r}_n^{surr})\right)e^{-\mathrm{s}\left[E_{n_{surr}}^{surr}(\mathbf{r}_n^{surr})+\mathscr{E}(\mathbf{r}_n^{QM},\mathbf{r}_n^{surr})\right]}}, \tag{58}$$

Equation (57) is in principle exact with any reasonable choice of $\mathscr{E}\left(\mathbf{r}_n^{QM},\mathbf{r}_n^{surr}\right)$, but fast numerical convergence is achieved if $\mathscr{E}\left(\mathbf{r}_n^{QM},\mathbf{r}_n^{surr}\right) \approx \Delta E_{n_{QM},n_{surr}}\left(\mathbf{r}_n^{QM},\mathbf{r}_n^{surr}\right)$. As is known from the literature on QM/MM modeling,[2,48] it may be difficult to choose a classical force field for a strongly QM part, or a QM part in which chemical transformations occur. However, the left hand side of Eq. (57) does not depend, given full sampling, on $\mathscr{E}\left(\mathbf{r}_n^{QM},\mathbf{r}_n^{surr}\right)$, and hence the results of QM/CG-MM computations based on Eq. (57), in contrast to subtractive QM/MM schemes,[2,48] will not be affected by this issue of using a classical force field for a quantum system, as soon as it provides a reasonable displacement of the surroundings molecules from the QM part.

**D. Possible strategies to approximate the electrostatic interactions between the QM part and the surroundings.**

By analogy with the well-known result from the theory of intermolecular interactions,[25] the electrostatic interaction energy between the QM part and the surroundings can be written as

$$\begin{aligned}\Delta E_{n_{QM},n_{surr}}^{(\text{elstat})}\left(\mathbf{r}_n^{QM},\mathbf{r}_n^{surr}\right) &= \langle n_{QM}^{(0)} n_{surr}^{(0)} | V_{QM-surr}(\mathbf{r}_e^{QM},\mathbf{r}_n^{QM},\mathbf{r}_e^{surr},\mathbf{r}_n^{surr}) | n_{QM}^{(0)} n_{surr}^{(0)} \rangle \\ &= \langle n_{QM}^{(0)} | V_{ext}^{elstat,n_{surr}}(\mathbf{r}_e^{QM},\mathbf{r}_n^{QM}) | n_{QM}^{(0)} \rangle,\end{aligned} \tag{59}$$

where $V_{ext}^{elstat,n_{surr}}$, defined by Eq. (36), is the external (from the viewpoint of the QM part) electrostatic potential created by the surroundings in the electronic state $n_{surr}$. Methods for calculating electrostatic interaction energies are well developed in computational chemistry. For example, Eq. (59) coincides, up to different notations, with the definition of $E^{Coul}$ in the effective fragment potential (EFP) method[32,34] implemented in GAMESS and other standard quantum chemical codes.[34] For this reason, we confine ourselves here to only two brief comments.

First, for the purpose of a rough estimate, one may represent each molecule in the surroundings as a set of multipoles (charge, dipole moment, quadrupole moment, etc.). In this case, if both the QM part and every molecule in the surroundings are electrically neutral, and contributions from the quadrupole and higher order multipoles are neglected, Eq. (59) simplifies to

$$\Delta E_{n_{QM},n_{surr}}^{(elstat)}\left(\mathbf{r}_n^{QM},\mathbf{r}_n^{surr}\right) = \sum_{i \in surr}\left(\frac{\left(\boldsymbol{\mu}^{QM}\cdot\boldsymbol{\mu}_i^{surr}\right)}{\left|(\mathbf{r}_n^{surr})_i\right|^3} - \frac{\left(\boldsymbol{\mu}^{QM}\cdot(\mathbf{r}_n^{surr})_i\right)\left((\mathbf{r}_n^{surr})_i\cdot\boldsymbol{\mu}_i^{surr}\right)}{\left|(\mathbf{r}_n^{surr})_i\right|^5}\right), \qquad (60)$$

where $(\mathbf{r}_n^{surr})_i$ are the coordinates of the *i*-th molecule in the surroundings (the QM part is supposed to be at the origin of the coordinates), $\boldsymbol{\mu}_i^{surr}$ is the dipole moment of the *i*-th molecule in the surroundings (including the induced dipoles caused by other molecules in the surroundings; note that the assumption of purely electrostatic interactions is introduced for the interactions between the QM part and the surroundings, but not within the QM part and not within the surroundings), and $\boldsymbol{\mu}^{QM}$ is the total dipole momentum of the QM part in vacuum for given nuclear coordinates:

$$\boldsymbol{\mu}^{QM}\left(\mathbf{r}_n^{QM}\right) = \left\langle n_{QM}^{(0)}\left|\sum_{j \in QM}(q_j\mathbf{r}_j^{QM})\right|n_{QM}^{(0)}\right\rangle, \qquad (61)$$

where index $j$ runs over all the electrons and the nuclei in the QM part, and $q_j$ and $\mathbf{r}_j^{QM}$ are charges and coordinates of the $j$-th particle, respectively.

Secondly, it is well known that one-center multipole expansions for the electrostatic interaction of molecules, such as Eq. (60), may not converge well with the multipoles order.[49] A better estimate of the electrostatic interaction energy can be obtained via the distributed multipolar analysis,[25,49] implemented, for example, in the EFP method.[32,34] In this approach, the electrostatic potential of a molecule is approximated by the Coulomb potential of a set of charges placed at atom positions and bond midpoints of the corresponding molecule. Then $\Delta E_{n_{QM},n_{surr}}^{(\text{elstat})}$ can be written as

$$\Delta E_{n_{QM},n_{surr}}^{(\text{elstat})}\left(\mathbf{r}_n^{QM},\mathbf{r}_n^{surr}\right) = \sum_{p_{QM}}\sum_{p_{surr}} \frac{q_{p_{QM}}(\mathbf{r}_n^{QM})q_{p_{surr}}(\mathbf{r}_n^{surr})}{\left|\mathbf{r}_{p_{QM}}(\mathbf{r}_n^{QM}) - \mathbf{r}_{p_{surr}}(\mathbf{r}_n^{surr})\right|}, \tag{62}$$

where the index $p_{QM}$ runs over all atoms and bond midpoints in the QM part, $p_{surr}$ runs over all atoms and bond midpoints in the surroundings, $q_{p_{QM}}$ and $q_{p_{surr}}$ are the effective charges placed on the corresponding atom or bond midpoint, and $\mathbf{r}_{p_{QM}}$ and $\mathbf{r}_{p_{surr}}$ are the corresponding coordinates. Computations by Eq. (62) are available in many quantum chemistry codes.

The use of an approximation given by Eq. (60) or (62), together with a specific choice of the mapping operators $\mathbf{M}_{\mathbf{R}}^N(\mathbf{r}_n^{surr})$, makes it possible to derive analytical expressions for the contributions to eigenvalues of the effective Hamiltonian $\hat{H}_{\text{eff}}^{QM}$ coming from the electrostatic interactions between the QM part and the surroundings, as shown in Sec. 4.1 of the Supplemental Material. In the cases when such analytical derivations cannot be performed, numerical techniques can be used, as discussed in Sec. 4.2 of the Supplemental Material.

**E. Schemes**

**Scheme 1.** Pseudo-algorithm for QM/CG-MM computations of the partition function *Z* and the expectation value of a QM operator $\langle \hat{O} \rangle_{QM/CG-MM}$ in the approximation of purely electrostatic interactions between the QM part and the surroundings (on steps 10, 11, 12, 13, 15, 16, choose variant "a"). Computations for a QM/CG-MM models including electrostatic and induction (corresponding to polarization of the QM part by the surroundings) interactions are also possible with this pseudo-algorithm (on the above-listed steps, choose variant "b").

**Step 0**. Computations for the isolated parts of the system.

    **Step 1**. Choose a set of geometries of the QM part $S_{QM}$ to be studied.

    **Step 2**. For each $\mathbf{r}_n^{QM} \in S_{QM}$ do:

        **Step 3**. Solve the Schrödinger equation for the QM part in vacuum using a standard quantum chemical code and record the resulting energies $E_{n_{QM}}^{QM}\left(\mathbf{r}_n^{QM}\right)$, wave functions $\Psi_{n_{QM}}^{QM}\left(\mathbf{r}_e^{QM};\mathbf{r}_n^{QM}\right)$ and, if required, derivatives of the energies $\partial^2 E_{n_{QM}}^{QM}\left(\mathbf{r}_n^{QM}\right) / \partial\left(\mathbf{r}_n^{QM}\right)_j \partial\left(\mathbf{r}_n^{QM}\right)_k$.

    **Step 4**. If available, obtain a CG model of the surroundings [including the expression for $V^{cl,surr}\left(\mathbf{r}_n^{surr}\right)$] from a library; otherwise, build a CG model of the surroundings from scratch, e.g. from classical MD simulations.

**Step 5**. Choose $\mathscr{E}\left(\mathbf{r}_n^{QM}, \mathbf{r}_n^{surr}\right)$, a computationally cheap approximation for the fine-grained energy of interaction between the QM part and the surroundings.

**Step 6**. Choose a set of the values of the CG variables $S_{surr}$ to be studied.

**Step 7**. For each $\mathbf{R}^N \in S_{surr}$ and for each $\mathbf{r}_n^{QM} \in S_{QM}$ and for each $n_{QM}$ in the set of the energy levels of the QM part do:

**Step 8**. Generate a set of $N_{sample}$ all-atom configurations of the system with the fixed geometry of the QM part $\mathbf{r}_n^{QM}$ and the distribution density for $\mathbf{r}_n^{surr}$ proportional to

$$u\left(\mathbf{R}^N - M^N(\mathbf{r}_n^{surr})\right) e^{-s\left[V^{cl,surr}(\mathbf{r}_n^{surr}) + \mathcal{E}(\mathbf{r}_n^{QM}, \mathbf{r}_n^{surr})\right]}.$$

**Step 9**. For $i_{sample} = 0, ..., N_{sample} - 1$ do:

**Step 10a**. Compute $\Delta E_{n_{QM},0}\left(\mathbf{r}_n^{QM}, \mathbf{r}_n^{surr}\right)$ using Eq. (60) [in the multipole approximation] or Eq. (62) [in the case of distributed multipolar analysis].

**Step 10b**. Compute $\Delta E_{n_{QM},0}\left(\mathbf{r}_n^{QM}, \mathbf{r}_n^{surr}\right)$ using Eqs. (60) or (62) of the main text and Eq. (S49) of the Supplemental Material.

**Step 11a**. Compute $f_1, f_2$, using Eq. (S28) of the Supplemental Material and, if required, $F_{1,jk}, F_{2,jk}, \mathbf{f}_j$ using Eq. (S32) of the Supplemental Material.

**Step 11b**. Compute $f_1$ and $f_2$, using Eq. (S28) of the Supplemental Material and $\Delta E_{n_{QM},0}$ from Step 10b.

**Step 12a**. Use the current values of $f_1, f_2, F_{1,jk}, F_{2,jk}, \mathbf{f}_j$ to compute the contribution of the $i_{sample}$–th frame to the averages in Eqs. (S27) and, if required, (S31) of the Supplemental Material.

**Step 12b**. Use the current values of $f_1$ and $f_2$ to compute the contribution of the $i_{sample}$–th frame to the averages in Eq. (S27) of the Supplemental Material.

**Step 13a**. Return the values of $E_{\text{eff},n_{QM}}\left(\mathbf{r}_n^{QM},\mathbf{R}^N\right)$ computed by Eq. (24) of the main text and Eq. (S27) of the Supplemental Material at Steps 9-12.

**Step 13b**. Return the values of $E_{\text{eff},n_{QM}}\left(\mathbf{r}_n^{QM},\mathbf{R}^N\right)$ computed by Eq. (39) of the main text and Eq. (S27) of the Supplemental Material at Steps 9-12.

**Step 14a**. If necessary, return the values of $\partial^2 E_{\text{eff},n_{QM}}^{(\text{elstat})}\left(\mathbf{r}_n^{QM},\mathbf{R}^N\right)\big/\partial\left(\mathbf{r}_n^{QM}\right)_j\partial\left(\mathbf{r}_n^{QM}\right)_k$ computed by Eq. (S31) of the Supplemental Material at Steps 9-12.

**Step 15a**. If required, compute and return the expectation value of a given operator $\hat{O}$ by Eq. (27).

**Step 15b**. If required, compute and return the expectation value of a given operator $\hat{O}$ by Eq. (42).

**Step 16a**. If required, compute and return the partition function of the system by Eq. (28), with $E_{\text{eff},n_{QM}} = E_{\text{eff},n_{QM}}^{(\text{elstat})}$.

**Step 16b**. If required, compute and return the partition function of the system by the analogue of Eq. (28), with $E_{\text{eff},n_{QM}} = E_{\text{eff},n_{QM}}^{(\text{elstat}+\text{ind QM})}$.

**Step 17**. If required, compute and return the expectation value of a given function $O_{\text{surr}}$ by Eqs. (14) and (15).

**Scheme 2.** Pseudo-algorithm for QM/CG-MM computations of the partition function $Z$ and the expectation value of a QM operator $\langle \hat{O} \rangle_{QM/CG-MM}$ in the approximation of electrostatic, induction and dispersion interactions between the QM part and the surroundings. If dispersion interactions are neglected, steps 11a and 13a should be omitted.

**Step 0**. Computations for the isolated parts of the system.

    **Step 1**. Choose a set of geometries of the QM part $S_{QM}$ to be studied.

    **Step 2**. For each $\mathbf{r}_n^{QM} \in S_{QM}$ do:

        **Step 3**. Solve the Schrödinger equation for the QM part in vacuum using a standard quantum chemical code and record the resulting energies $E_{n_{QM}}^{QM}(\mathbf{r}_n^{QM})$ and wave functions $\Psi_{n_{QM}}^{QM}(\mathbf{r}_e^{QM}; \mathbf{r}_n^{QM})$.

    **Step 4**. If available, obtain a CG model of the surroundings [including the expression for $V^{cl,surr}(\mathbf{r}_n^{surr})$] from a library; otherwise, build a CG model of the surroundings from scratch, e.g. from classical MD simulations.

    **Step 5**. If available, obtain polarizabilities $r_{rs}^{surr,I,n_{surr}}$ and/or parameters $s_{rs}^{surr,I,n_{surr}}$ for each CG site from a library; otherwise, run QM computations for each molecule from the surroundings in vacuum and compute $r_{rs}^{surr,I,n_{surr}}$ and/or $s_{rs}^{surr,I,n_{surr}}$ by Eqs. (37) and (38). If required, do the same for higher multipole polarizabilities, hyperpolarizabilitites, etc.

**Step 6**. Choose $\mathscr{E}(\mathbf{r}_n^{QM}, \mathbf{r}_n^{surr})$, a computationally cheap approximation for the fine-grained energy of interaction between the QM part and the surroundings.

**Step 7**. Choose a set of the values of the CG variables $S_{surr}$ to be studied.

**Step 8**. For each $\mathbf{R}^N \in S_{surr}$ and for each $\mathbf{r}_n^{QM} \in S_{QM}$ and for each $n_{QM}$ in the set of the energy levels of the QM part do:

> **Step 9**. Generate a set of $N_{sample}$ all-atom configurations of the system with the fixed geometry of the QM part $\mathbf{r}_n^{QM}$ and the distribution density for $\mathbf{r}_n^{surr}$ proportional to
>
> $$\mathsf{u}\left(\mathbf{R}^N - M^N(\mathbf{r}_n^{surr})\right) e^{-\mathsf{s}\left[V^{cl,surr}(\mathbf{r}_n^{surr}) + \mathcal{E}(\mathbf{r}_n^{QM}, \mathbf{r}_n^{surr})\right]}.$$
>
> **Step 10**. For $i_{sample} = 0, ..., N_{sample} - 1$ do:
>
> > **Step 11**. Compute the electrostatic and induction contributions to $\Delta E_{n_{QM},0}^{(elstat+ind)}\left(\mathbf{r}_n^{QM}, \mathbf{r}_n^{surr}\right)$ by Eq. (35).
> >
> > **Step 11a**. Compute the dispersion contribution to $\Delta E_{n_{QM},0}^{(disp)}\left(\mathbf{r}_n^{QM}, \mathbf{r}_n^{surr}\right)$ by Eq. (44).
> >
> > **Step 12**. Compute $f_1$ and $f_2$, using Eq. (S28) of the Supplemental Material.
> >
> > **Step 13**. Compute the induction contribution to the coefficients $k_{i,j}^{(ind)n_{QM},n_{surr}}$ using Eqs. (33), (34) of the main text and Eq. (S34) of the Supplemental Material.
> >
> > **Step 13a**. Compute the dispersion contribution to the coefficients $k_{i,j}^{(disp)n_{QM},n_{surr}}$ using Eq. (45) of the main text and Eq. (S35) of the Supplemental Material.
> >
> > **Step 14**. Use the current values of $k_{i,j}^{n_{QM},n_{surr}}$, $f_1$ and $f_2$ to compute the contribution of the $i_{sample}$–th frame to the averages included in Eq. (19).
>
> **Step 15**. Use the averages computed at Steps 10-14 to compute the matrix
>
> $\left\langle i^{(0)} \left| e^{-\mathsf{s}\hat{H}_{eff}^{QM}} \right| j^{(0)} \right\rangle$ by Eq. (19).
>
> **Step 16**. Compute eigenvalues and eigenfunctions of the effective QM Hamiltonian by diagonalization of the matrix $\left\langle i^{(0)} \left| e^{-\mathsf{s}\hat{H}_{eff}^{QM}} \right| j^{(0)} \right\rangle$ computed at Step 15.

**Step 17**. If required, compute and return the expectation value of a given operator $\hat{O}$ by Eq. (42), with $E_{\mathit{eff},n_{QM}} = E_{\mathit{eff},n_{QM}}^{(\mathit{elstat}+\mathit{ind}+\mathit{disp})}$.

**Step 18**. If required, compute and return the partition function of the system by Eq. (28), with $E_{\mathit{eff},n_{QM}} = E_{\mathit{eff},n_{QM}}^{(\mathit{elstat}+\mathit{ind}+\mathit{disp})}$.

**Step 19**. If required, compute and return the expectation value of a given function $O_{\mathit{surr}}$ by Eqs. (14) and (15).

# Quantum Mechanics / Coarse-Grained Molecular Mechanics (QM/CG-MM)

Anton V. Sinitskiy and Gregory A. Voth[1a)]

[1] Department of Chemistry, James Franck Institute, and Institute for Biophysical Dynamics, The University of Chicago, Chicago, Illinois 60637, United States

[a)] Author to whom correspondence should be addressed: gavoth@uchicago.edu

**SUPPLEMENTARY MATERIAL**

## 1.    Derivation of the effective Hamiltonian of the QM part

In this section, we derive an explicit expression for the effective QM Hamiltonian $\hat{H}_{eff}^{QM}$ that ensures the consistency condition, Eq. (7). By changing the order of integration, Eq. (5) can be transformed to

$$\langle \hat{O} \rangle_{QM/CG-MM} = \frac{1}{Z_{QM/CG}} \left( \int \frac{d\mathbf{p}_n^{QM}}{(2f\hbar)^{3N_n^{QM}}} e^{-\mathsf{S}T_n^{QM}(\mathbf{p}_n^{QM})} \right) \left( \int \frac{d\mathbf{P}^N}{(2f\hbar)^{3N}} e^{-\mathsf{S}T_{CG}(\mathbf{P}^N)} \right) \\ \times \int d\mathbf{r}_n^{QM} d\mathbf{R}^N \, \mathrm{Tr}_{r_e^{QM}} \left\{ e^{-\mathsf{S}\hat{H}_{eff}^{QM}} \hat{O} \right\}. \qquad (S63)$$

On the other hand, plugging in the following identity:

$$1 = \int d\mathbf{R}^N \mathsf{u}\left(\mathbf{R}^N - \mathbf{M}_{\mathbf{R}}^N(\mathbf{r}_n^{surr})\right), \qquad (S64)$$

where the mapping operators $\mathbf{M}_{\mathbf{R}_I}(\mathbf{r}_n^{surr})$ and the CG variables $\mathbf{R}^N = \{\mathbf{R}_1, \ldots, \mathbf{R}_N\}$ were introduced in the main text, into the right hand side of Eq. (2), the following expression can be obtained:

$$\langle \hat{O} \rangle = \frac{1}{Z} \left( \int \frac{d\mathbf{p}_n}{(2f\hbar)^{3N_n^{sys}}} e^{-\mathsf{S}T_n(\mathbf{p}_n)} \right) \cdot \int d\mathbf{r}_n^{QM} d\mathbf{R}^N d\mathbf{r}_n^{surr} \mathsf{u}\left(\mathbf{R}^N - \mathbf{M}_{\mathbf{R}}^N(\mathbf{r}_n^{surr})\right) \mathrm{Tr}_{r_e} \left\{ e^{-\mathsf{S}\hat{H}_e} \hat{O} \right\}. \quad (S65)$$

Comparison of the right hand sides of Eqs. (S63) and (S65) demonstrates that they have a similar structure: prefactors independent of $\mathbf{r}_e^{QM}$, $\mathbf{r}_n^{QM}$ and $\mathbf{R}^N$ are followed by integrals over $\mathbf{r}_n^{QM}$ and $\mathbf{R}^N$ of some expressions involving traces over electronic degrees of freedom.

It is instructive to consider first how the consistency condition could be satisfied if the electrons were distinguishable particles. Then the traces in Eqs. (S63) and (S65) could be simply written in the local bases as integrals of the corresponding diagonal matrix elements over all the coordinates of the electrons in the system $N_e^{sys}$, or over the first $N_e^{QM}$ coordinates, respectively:

$$\mathrm{Tr}_{r_e}^{distinguishable}\left\{e^{-s\hat{H}_e}\hat{O}\right\} = \int d\mathbf{r}_e^{QM} d\mathbf{r}_e^{surr} \left\langle \mathbf{r}_e^{QM}, \mathbf{r}_e^{surr} \left| e^{-s\hat{H}_e}\hat{O} \right| \mathbf{r}_e^{QM}, \mathbf{r}_e^{surr} \right\rangle, \quad (S66)$$

$$\mathrm{Tr}_{r_e^{QM}}^{distinguishable}\left\{e^{-s\hat{H}_{eff}^{QM}}\hat{O}\right\} = \int d\mathbf{r}_e^{QM} \left\langle \mathbf{r}_e^{QM} \left| e^{-s\hat{H}_{eff}^{QM}}\hat{O} \right| \mathbf{r}_e^{QM} \right\rangle, \quad (S67)$$

and the consistency condition would be fulfilled with the following definition of the effective QM Hamiltonian [identical to Eq. (8) in the main text]:

$$e^{-s\hat{H}_{eff}^{QM}} = C \cdot \int d\mathbf{r}_n^{surr} \mathsf{u}\left(\mathbf{R}^N - \mathbf{M}_\mathbf{R}^N(\mathbf{r}_n^{surr})\right)\mathrm{Tr}_{r_e^{surr}}\left\{e^{-s\hat{H}_e}\right\}, \quad (S68)$$

where $C$ is a constant determined by the prefactors on the right hand sides of Eqs. (S63) and (S65), and the trace is interpreted as an integral over the last $N_e^{surr} = N_e^{sys} - N_e^{QM}$ coordinates:

$$\mathrm{Tr}_{r_e^{surr}}^{distinguishable}\left\{e^{-s\hat{H}_e}\right\} = \int d\mathbf{r}_e^{surr} \left\langle \mathbf{r}_e^{surr} \left| e^{-s\hat{H}_e} \right| \mathbf{r}_e^{surr} \right\rangle. \quad (S69)$$

However, all the electrons in the system are in fact indistinguishable from each other. To write the expressions for the traces, one must use only basis functions corresponding to physically admissible states.[1] For example, in the local basis,

$$\{_{basis}^{sys}(\mathbf{r}_e^{QM}, \mathbf{r}_e^{surr}) = \sqrt{(N_e^{sys}!)}\,\mathcal{A}_{sys}\left(\left|\mathbf{r}_e^{QM}\right\rangle \left|\mathbf{r}_e^{surr}\right\rangle\right), \quad (S70)$$

$$\{_{basis}^{QM}(\mathbf{r}_e^{QM}) = \sqrt{(N_e^{QM}!)}\,\mathcal{A}_{QM}\left(\left|\mathbf{r}_e^{QM}\right\rangle\right), \quad (S71)$$

where $\mathcal{A}_{sys}$ is the antisymmetrization operator for all the electrons in the system, $\mathcal{A}_{QM}$ is the antisymmetrization operator for the electrons in the QM part, and the prefactors $\sqrt{(N_e^{sys}!)}$ and $\sqrt{(N_e^{QM}!)}$ are included to ensure the unit normalization of the antisymmetrized basis functions. Then the traces in Eqs. (S63) and (S65) are naturally interpreted as

$$\mathrm{Tr}_{r_e}\left\{e^{-s\hat{H}_e}\hat{O}\right\} = (N_e^{sys}!)\int d\mathbf{r}_e^{QM} d\mathbf{r}_e^{surr} \left\langle \mathbf{r}_e^{QM}, \mathbf{r}_e^{surr} \middle| \mathcal{A}_{sys} e^{-s\hat{H}_e} \hat{O} \mathcal{A}_{sys} \middle| \mathbf{r}_e^{QM}, \mathbf{r}_e^{surr} \right\rangle, \quad (S72)$$

$$\mathrm{Tr}_{r_e^{QM}}\left\{e^{-s\hat{H}_{eff}^{QM}}\hat{O}\right\} = (N_e^{QM}!)\int d\mathbf{r}_e^{QM} \left\langle \mathbf{r}_e^{QM} \middle| \mathcal{A}_{QM} e^{-s\hat{H}_{eff}^{QM}} \hat{O} \mathcal{A}_{QM} \middle| \mathbf{r}_e^{QM} \right\rangle. \quad (S73)$$

It turns out that, in this case, the expression $\int d\mathbf{r}_n^{QM} d\mathbf{R}^N \mathrm{Tr}_{r_e^{QM}}\left\{e^{-s\hat{H}_{eff}^{QM}}\right\}$ can be transformed to the form of $\int d\mathbf{r}_n^{QM} d\mathbf{r}_n^{surr} \mathrm{Tr}_{r_e}\left\{e^{-s\hat{H}_e}\right\}$, hence the full partition function of the system $Z$ can be related to the partition function $Z_{QM/CG-MM}$, naturally defined in the QM/CG-MM approach by Eq. (6), if the effective QM Hamiltonian is defined by the same Eq. (8), but with the following natural interpretation of the trace over the positions of the electrons in the surroundings [compare to Eq. (S69) and bear in mind Eqs. (S70) and (S71)]:

$$\mathrm{Tr}_{r_e^{surr}}\left\{e^{-s\hat{H}_e}\right\} = \int d\mathbf{r}_e^{QM} d\mathbf{r}_e'^{QM} d\mathbf{r}_e^{surr} \left(\sqrt{(N_e^{QM}!)} \mathcal{A}_{QM} \middle| \mathbf{r}_e^{QM} \right\rangle)$$
$$\times \left\langle \mathbf{r}_e^{QM}, \mathbf{r}_e^{surr} \middle| \sqrt{(N_e^{sys}!)} \mathcal{A}_{sys} e^{-s\hat{H}_e} \sqrt{(N_e^{sys}!)} \mathcal{A}_{sys} \middle| \mathbf{r}_e'^{QM}, \mathbf{r}_e^{surr} \right\rangle \left(\sqrt{(N_e^{QM}!)} \left\langle \mathbf{r}_e'^{QM} \middle| \mathcal{A}_{QM} \right)$$
$$(S74)$$

As follows from Eqs. (1), (6), (8), (10), (S73) and (S74), the explicit expression for the constant $C$ is

$$C = \frac{Z_{QM/CG-MM}}{Z} \frac{1}{\left(N_e^{QM}!\right)^2} \frac{\left(\int \frac{d\mathbf{p}_n^{surr}}{(2f\hbar)^{3N_n^{surr}}} e^{-\mathrm{S}T_n^{surr}(\mathbf{p}_n^{surr})}\right)}{\left(\int \frac{d\mathbf{P}^N}{(2f\hbar)^{3N}} e^{-\mathrm{S}T_{CG}(\mathbf{P}^N)}\right)}. \quad (S75)$$

One may require that the QM/CG-MM model give the exact result for the partition function $Z$, that is $Z_{QM/CG-MM} = Z$; then Eq. (12) will define the constant $C$ that should be used in Eq. (8). This variant is used in the main text of the article, Eq. (12). Alternatively, one may instead choose a specific expression for $C$ for some other reasons of convenience. In this case, $Z$ and $Z_{QM/CG-MM}$ will not coincide, but one will be able to easily find $Z$ using Eq. (12) and the preselected definition of $C$, as soon as $Z_{QM/CG-MM}$ is computed from the QM/CG-MM model.

It turns out that the more general expression $\int d\mathbf{r}_n^{QM} d\mathbf{R}^N \, \mathrm{Tr}_{r_e^{QM}} \left\{ e^{-\mathrm{S}\hat{H}_{eff}^{QM}} \hat{O} \right\}$ cannot be similarly transformed to the form of $\int d\mathbf{r}_n^{QM} d\mathbf{r}_n^{surr} \, \mathrm{Tr}_{r_e} \left\{ e^{-\mathrm{S}\hat{H}_e} \hat{O} \right\}$ in the case of an arbitrary operator $\hat{O}$, if the definition of the effective QM Hamiltonian given by Eq. (8) is used, and therefore condition (7), strictly speaking, is not satisfied in the general case. However, we claim that the definition of the effective QM Hamiltonian given by Eq. (8), with the interpretation of the trace given by Eq. (S74) and the value of $C$ given by Eq. (12), should be used in the QM/CG-MM approach. The physical motivation for this claim is given in Appendix B in the main text of the article.

2. **Derivation of the probability distribution for the CG variables**

By inserting identity (S64) into the integrand in Eq. (13), we obtain an expression that can be rewritten as

$$\langle O_{surr} \rangle = \int d\mathbf{R}^N p_{surr}(\mathbf{R}^N) O_{surr}(\mathbf{R}^N), \tag{S76}$$

where $p_{surr}(\mathbf{R}^N)$ is the function defined by

$$p_{surr}(\mathbf{R}^N) \propto \int d\mathbf{r}_n \mathsf{u}(\mathbf{R}^N - M^N(\mathbf{r}_n^{surr})) \mathrm{Tr}_{r_e}\{e^{-\mathrm{s}\hat{H}_e}\}, \qquad \int d\mathbf{R}^N p_{surr}(\mathbf{R}^N) = 1. \tag{S77}$$

As claimed in the main text, $p_{surr}(\mathbf{R}^N)$ can be interpreted as the probability distribution for the CG variables.

Our goal now is to explicitly describe the influence of the QM part on the surroundings by means of expressing $p_{surr}(\mathbf{R}^N)$ in terms of the properties of the QM region. To do so, we notice that the trace of the exponential of the effective QM Hamiltonian over the positions of the electrons in the QM part can be related to the full trace of the exponential of the full electron Hamiltonian of the system, as follows from Eqs. (8), (S73) and (S74):

$$\mathrm{Tr}_{r_e^{QM}}\{e^{-\mathrm{s}\hat{H}_{eff}^{QM}}\} = C(N_e^{QM}!)^2 \int d\mathbf{r}_n^{surr} \mathsf{u}(\mathbf{R}^N - \mathbf{M}_{\mathbf{R}}^N(\mathbf{r}_n^{surr})) \mathrm{Tr}_{r_e}\{e^{-\mathrm{s}\hat{H}_e}\}, \tag{S78}$$

and therefore, according to Eq. (S77),

$$p_{surr}(\mathbf{R}^N) \propto \int d\mathbf{r}_n^{QM} \mathrm{Tr}_{r_e^{QM}}\{e^{-\mathrm{s}\hat{H}_{eff}^{QM}}\}. \tag{S79}$$

Taking the trace in the basis of the eigenfunctions of the effective QM Hamiltonian, we come to Eq. (15) from the main text.

To extend the analogy with the MS-CG method mentioned in the main text, one can define a coarse-grained potential $V_{CG}(\mathbf{R}^N)$ such that the probability distribution for the CG variables follows the Boltzmann law for this potential:[2]

$$p_{surr}(\mathbf{R}^N) = C'e^{-\mathrm{s}V_{CG}(\mathbf{R}^N)}, \tag{S80}$$

where $C'$ is a normalization constant. Then

$$V_{CG}(\mathbf{R}^N) = -\frac{1}{S}\ln\left[\int d\mathbf{r}_n^{QM} \sum_i e^{-S E_{eff}^i(\mathbf{r}_n^{QM},\mathbf{R}^N)}\right] + const. \qquad (S81)$$

Notice that if $E_{eff,i}(\mathbf{r}_n^{QM},\mathbf{R}^N)$ as functions of $\mathbf{r}_n^{QM}$ can be closely approximated by a second order Taylor series expansion, for example if the QM part includes only one relatively rigid molecule, then the integral over $d\mathbf{r}_n^{QM}$ in Eq. (15) can be computed analytically, leading to

$$p_{surr}(\mathbf{R}^N) \propto \sum_i \left[\frac{e^{-S E_{eff,i}(\mathbf{r}_n^{QM,eq},\mathbf{R}^N)}}{\prod_m \check{S}_{i,m}^{QM,eq}(\mathbf{R}^N)}\right], \qquad (S82)$$

where $\mathbf{r}_n^{QM,eq}$ are the equilibrium values of the coordinates of the nuclei in the QM part and $\check{S}_{i,m}^{QM,eq}(\mathbf{R}^N)$ are the normal mode frequencies for the *i*-th electronic state of the QM part at given values of CG variables $\mathbf{R}^N$.

3. **Derivation of the matrix representation of the effective Hamiltonian**

The exponential of the full electronic Hamiltonian can be written via a spectral representation:

$$e^{-S\hat{H}_e} = \sum_{n_{QM},n_{surr}} |n_{QM} n_{surr}\rangle \frac{e^{-S E_{n_{QM},n_{surr}}^{sys}}}{\langle n_{QM} n_{surr} | n_{QM} n_{surr}\rangle} \langle n_{QM} n_{surr}|, \qquad (S83)$$

where $|n_{QM} n_{surr}\rangle$ is a short-hand notation for the exact wave function of the system, $\Psi_{n_{QM},n_{surr}}^{sys}$. By plugging Eq. (S83) into Eq. (8) and putting the resulting operators between the bra- and ket-vectors $\langle i^{(0)}|$ and $|j^{(0)}\rangle$, the matrix elements of $e^{-S\hat{H}_{eff}^{QM}}$ in the basis of the wave functions of the QM part in vacuum can be expressed as

$$\left\langle i^{(0)}\left|e^{-s\hat{H}_{eff}^{QM}}\right|j^{(0)}\right\rangle = C\sum_{n_{QM},n_{surr}}\int d\mathbf{r}_n^{surr}\left[\mathsf{u}\left(\mathbf{R}^N - M^N(\mathbf{r}_n^{surr})\right)\right.$$

$$\left. \times e^{-sE_{n_{QM},n_{surr}}^{sys}}\frac{\left\langle i^{(0)}\left|\mathrm{Tr}_{r_e^{surr}}\left\{\left|n_{QM}n_{surr}\right\rangle\left\langle n_{QM}n_{surr}\right|\right\}\right|j^{(0)}\right\rangle}{\left\langle n_{QM}n_{surr}\left|n_{QM}n_{surr}\right\rangle}\right], \quad (S84)$$

where $\left|j^{(0)}\right\rangle$ is a short-hand notation for $\Psi_j^{QM}$. The superscript (0) denotes that these wave functions refer to the isolated ("unperturbed") QM part, unlike $\left|n_{QM}n_{surr}\right\rangle$ in Eq. (S83) that refer to the system with the interactions turned on. As a rough estimate, if one omits the coefficients $c_{k,l}^{n_{QM},n_{surr}}$ and permutations of electrons between the QM part and the surroundings in the antisymmetrization operator $\mathcal{A}_{sys}$ in Eq. (17), which physically corresponds to neglecting exchange interactions between the QM part and the surroundings and neglecting corrections to the wave functions caused by the polarization, then the following estimate can be obtained:

$$\frac{\left\langle i^{(0)}\left|\mathrm{Tr}_{r_e^{surr}}\left\{\left|n_{QM}n_{surr}\right\rangle\left\langle n_{QM}n_{surr}\right|\right\}\right|j^{(0)}\right\rangle}{\left\langle n_{QM}n_{surr}\left|n_{QM}n_{surr}\right\rangle} \approx \mathsf{u}_{i,n_{QM}}\mathsf{u}_{j,n_{QM}}. \quad (S85)$$

This motivates introducing the coefficients $k_{i,j}^{n_{QM},n_{surr}}$ defined in the main text, Eq. (22), as small parameters.

With the use of Eqs. (18) and (22), the expression for the matrix elements of $e^{-s\hat{H}_{eff}^{QM}}$ assumes the following form:

$$\left\langle i^{(0)}\left|e^{-s\hat{H}_{eff}^{QM}}\right|j^{(0)}\right\rangle = C\sum_{n_{QM}}e^{-sE_{n_{QM}}^{QM}(\mathbf{r}_n^{QM})}\sum_{n_{surr}}\int d\mathbf{r}_n^{surr}\left[\mathsf{u}\left(\mathbf{R}^N - \mathbf{M}_\mathbf{R}^N(\mathbf{r}_n^{surr})\right)e^{-sE_{n_{surr}}^{surr}(\mathbf{r}_n^{surr})}\right.$$

$$\left. \times\left(\mathsf{u}_{i,j}\mathsf{u}_{i,n_{QM}} + k_{i,j}^{n_{QM},n_{surr}}(\mathbf{r}_n^{QM},\mathbf{r}_n^{surr})\right)e^{-s\Delta E_{n_{QM},n_{surr}}(\mathbf{r}_n^{QM},\mathbf{r}_n^{surr})}\right]. \quad (S86)$$

In the main text, we introduced averaging $\langle\cdots\rangle_{\mathbf{R}^N}$, Eq. (21), and the CG potential for the surroundings in the absence of the QM part $V_{CG}^{surr}(\mathbf{R}^N)$, Eq. (20). These two definitions are similar to ones in the MS-CG theory [see Eqs. (20) and (28) in ref. [2]], but are more general in that they allow for excited electronic states of the surroundings, which is reflected by the dependence of the fine-grained energy $E^{surr}_{n_{surr}}$ and the function to be averaged $f$ on the index $n_{surr}$, and by the summation over this index in both definitions. With the use of these two notations, Eq. (S86) transforms to the form given in the main text, Eq. (19).

## 4. Technical aspects of averaging the energy of electrostatic interactions between the QM part and the surroundings

### 4.1. Analytical averaging

In some cases, averaging $e^{-s\Delta E_{n_{QM},n_{surr}}}$ over the configurations of the surroundings compatible with the given values of CG variables $\mathbf{R}^N$ can be performed analytically. For example, if the CG variables $\mathbf{R}^N = \{\mathbf{R}^N_c, \Omega^N\}$ are defined as the centers of mass $\mathbf{R}^N_c$ and the rotation angles $\Omega^N$ of each of $N$ molecules in the surroundings, and if the molecules in the surroundings are approximated by rigid bodies, then the coordinates of each nuclei in the surroundings $\mathbf{r}^{surr}_n$ can be uniquely determined using the values of $\mathbf{R}^N$, and therefore $\Delta E^{(elstat)}_{n_{QM},n_{surr}}$ can be represented as an entity depending on $\mathbf{r}^{surr}_n$ only via the mapping operators: $\Delta E^{(elstat)}_{n_{QM},n_{surr}}\left(\mathbf{r}^{QM}_n, \mathbf{r}^{surr}_n\right) = \Delta E^{(elstat)}_{n_{QM},n_{surr}}\left(\mathbf{r}^{QM}_n, \mathbf{M}^N_\mathbf{R}(\mathbf{r}^{surr}_n)\right)$. Then, Eq. (25) simplifies to

$$\Delta E^{(elstat)}_{eff,n_{QM}}\left(\mathbf{r}^{QM}_n, \mathbf{R}^N\right) = \Delta E^{(elstat)}_{n_{QM},n_{surr}}\left(\mathbf{r}^{QM}_n, \mathbf{r}^{surr}_n(\mathbf{R}^N)\right). \tag{S87}$$

In particular, in the multipole approximation, Eq. (60), the electrostatic contribution to the $n_{QM}$-th effective QM energy level can be written as

$$\Delta E_{\text{eff},n_{QM}}^{\text{(elstat)}}\left(\mathbf{r}_n^{QM},\mathbf{R}^N\right) = \sum_{I=1}^{N}\left(\frac{\left(\boldsymbol{\mu}^{QM}\cdot\boldsymbol{\mu}_I^{CG}(\Omega_I)\right)}{\left|\mathbf{R}_{c,I}\right|^3} - \frac{\left(\boldsymbol{\mu}^{QM}\cdot\mathbf{R}_{c,I}\right)\left(\mathbf{R}_{c,I}\cdot\boldsymbol{\mu}_I^{CG}(\Omega_I)\right)}{\left|\mathbf{R}_{c,I}\right|^5}\right), \tag{S88}$$

where $I$ enumerates the CG sites (molecules in the surroundings), $\mathbf{R}_{c,I}$ and $\Omega_I$ are the center of mass and rotation angles of the $I$-th CG site, respectively, and $\boldsymbol{\mu}_I^{CG}$ is the dipole moment of the $I$-th CG site (its direction depends on the rotation angles $\Omega_I$). A similar derivation can be performed in the case of the distributed multipolar analysis, Eq. (62), since the positions of the effective charges in the surroundings in the approximation of rigid molecules can also be uniquely determined from the values of the CG variables. Note that the mapping operators for $\Omega^N$ are nonlinear functions of the nuclear coordinates $\mathbf{r}_n^{\text{surr}}$. In the original MS-CG method, all mapping operators were assumed to be linear, which is necessary for force-matching to work.[2,3] However, in the present paper the condition of linearity of $\mathbf{M}_{\mathbf{R}}^N$ has not been used, allowing for the use of nonlinear CG variables, such as orientation angles.

## *4.2. Numerical averaging*

Numerical averaging of $e^{-s\Delta E_{n_{QM},n_{\text{surr}}}}$ is a powerful alternative to analytical approaches that is not restricted to specific choices of the CG variables. As discussed in Appendix C, it may be efficient to use a computationally inexpensive approximation for the interaction energy $\mathcal{E}\left(\mathbf{r}_n^{QM},\mathbf{r}_n^{\text{surr}}\right)$ to get the desired average as

$$\Delta E_{\text{eff},i}^{\text{(elstat)}}\left(\mathbf{r}_n^{QM},\mathbf{R}^N\right) = -\frac{1}{s}\ln\langle f_1\rangle_{\mathbf{R}^N,ac} + \frac{1}{s}\ln\langle f_2\rangle_{\mathbf{R}^N,ac}, \tag{S89}$$

where

$$f_1 = e^{-s\left[\Delta E_{i,\text{surr}}(\mathbf{r}_n^{QM},\mathbf{r}_n^{\text{surr}})-\mathcal{E}(\mathbf{r}_n^{QM},\mathbf{r}_n^{\text{surr}})\right]}, \qquad f_2 = e^{s\mathcal{E}(\mathbf{r}_n^{QM},\mathbf{r}_n^{\text{surr}})}. \tag{S90}$$

To do so, one may generate an ensemble at the atomistic resolution level with the distribution density proportional to $\mathsf{u}\left(\mathbf{R}^N - \mathbf{M}_\mathbf{R}^N(\mathbf{r}_n^{surr})\right) e^{-\mathsf{s}\left[V^{cl,surr}(\mathbf{r}_n^{surr}) + \mathcal{E}(\mathbf{r}_n^{QM}, \mathbf{r}_n^{surr})\right]}$, for example, by biased classical molecular dynamics (MD) or Monte-Carlo (MC) simulations. Here $V^{cl,surr}\left(\mathbf{r}_n^{surr}\right)$ is a classical force field approximating the value of $-\mathsf{s}^{-1} \ln \sum_{n_{surr}} e^{-\mathsf{s} E_{n_{surr}}^{surr}(\mathbf{r}_n^{surr})}$. The delta function could, in practice, be approximated by a Gaussian function penalizing deviations of the actual values of $\mathbf{M}_\mathbf{R}^N(\mathbf{r}_n^{surr})$ from the target values $\mathbf{R}^N$, which corresponds to adding a harmonic restraint to the potential $V^{cl,surr} + \mathcal{E}$. After that, the values of $\langle f_1 \rangle_{\mathbf{R}^N,ac}$ and $\langle f_2 \rangle_{\mathbf{R}^N,ac}$ can be found simply as arithmetic averages of $f_1$ and $f_2$, respectively, over such an ensemble. Simultaneously with introducing a classical force field $V^{cl,surr}\left(\mathbf{r}_n^{surr}\right)$ for the surroundings, it makes sense to neglect electronic excitations in the surroundings. In the ensuing formulas within this subsection, it is assumed that only the value of $n_{surr} = 0$, corresponding to the ground state of the surroundings, is used.

An alternative way might be to run classical MD or MC simulations with the potential $V^{cl,surr} + \mathcal{E}$, such that various values of $\mathbf{R}^N$ are sampled along the trajectory; for each frame $t$, to compute the current values of $f_1(t), f_2(t)$ and $\mathbf{M}_\mathbf{R}^N\left(\mathbf{r}_n^{surr}(t)\right)$; after that, to find coefficients $c_1^i$ and $c_2^i$ in the approximations

$$\langle f_1 \rangle_{\mathbf{R}^N,ac} = \sum_i c_1^i e_1^i\left(\mathbf{R}^N\right), \qquad \langle f_2 \rangle_{\mathbf{R}^N,ac} = \sum_i c_2^i e_2^i\left(\mathbf{R}^N\right), \tag{S91}$$

where $e_1^i\left(\mathbf{R}^N\right)$ and $e_2^i\left(\mathbf{R}^N\right)$ are some basis functions, by the method of least squares:

$$\{c_1^i\} = \arg\min \sum_t \left( f_1(t) - \sum_i c_1^i e_1^i \left( \mathbf{M_R^N}\left( \mathbf{r}_n^{surr}(t) \right) \right) \right)^2,$$

$$\{c_2^i\} = \arg\min \sum_t \left( f_2(t) - \sum_i c_2^i e_2^i \left( \mathbf{M_R^N}\left( \mathbf{r}_n^{surr}(t) \right) \right) \right)^2.$$
(S92)

Analytical derivatives of the effective energy can be calculated in a similar way. For example, the elements of the Hessian matrix equal

$$\frac{\partial^2}{\partial \left(\mathbf{r}_n^{QM}\right)_j \partial \left(\mathbf{r}_n^{QM}\right)_k} E_{\text{eff},i}^{(\text{elstat})}\left(\mathbf{r}_n^{QM}, \mathbf{R}^N\right) = \frac{\partial^2}{\partial \left(\mathbf{r}_n^{QM}\right)_j \partial \left(\mathbf{r}_n^{QM}\right)_k} E_i^{QM}\left(\mathbf{r}_n^{QM}\right)$$
$$+ \frac{\langle F_{1,jk} \rangle_{\mathbf{R}^N,ac} - \mathsf{s}\langle F_{2,jk} \rangle_{\mathbf{R}^N,ac}}{\langle f_1 \rangle_{\mathbf{R}^N,ac}} + \mathsf{s}\frac{\langle \mathbf{f}_j \rangle_{\mathbf{R}^N,ac} \otimes \langle \mathbf{f}_k \rangle_{\mathbf{R}^N,ac}}{\left(\langle f_1 \rangle_{\mathbf{R}^N,ac}\right)^2}$$
(S93)

where

$$F_{1,jk} = \left[\frac{\partial^2 \Delta E_{n_{QM},n_{surr}}^{(\text{elstat})}(\mathbf{r}_n^{QM}, \mathbf{r}_n^{surr})}{\partial (\mathbf{r}_n^{QM})_j \partial (\mathbf{r}_n^{QM})_k}\right] e^{\mathsf{s}\mathcal{E}(\mathbf{r}_n^{QM}, \mathbf{r}_n^{surr})},$$

$$F_{2,jk} = \left[\frac{\partial \Delta E_{n_{QM},n_{surr}}^{(\text{elstat})}(\mathbf{r}_n^{QM}, \mathbf{r}_n^{surr})}{\partial (\mathbf{r}_n^{QM})_j} \otimes \frac{\partial \Delta E_{n_{QM},n_{surr}}^{(\text{elstat})}(\mathbf{r}_n^{QM}, \mathbf{r}_n^{surr})}{\partial (\mathbf{r}_n^{QM})_k}\right] e^{\mathsf{s}\mathcal{E}(\mathbf{r}_n^{QM}, \mathbf{r}_n^{surr})},$$
(S94)

$$\mathbf{f}_j = \frac{\partial \Delta E_{n_{QM},n_{surr}}^{(\text{elstat})}(\mathbf{r}_n^{QM}, \mathbf{r}_n^{surr})}{\partial (\mathbf{r}_n^{QM})_j} e^{\mathsf{s}\mathcal{E}(\mathbf{r}_n^{QM}, \mathbf{r}_n^{surr})}$$

and indices $j$ and $k$ run over all the nuclei in the QM part.

### 4.3. Small β expansion

Finally, note that in small $\beta$ expansion Eq. (25) can be rewritten in a simple form:

$$\Delta E_{\text{eff},n_{QM}}^{(\text{elstat})}\left(\mathbf{r}_n^{QM}, \mathbf{R}^N\right) = \langle \Delta E_{n_{QM},n_{surr}}^{(\text{elstat})}(\mathbf{r}_n^{QM}, \mathbf{r}_n^{surr}) \rangle_{\mathbf{R}^N}$$
$$- \mathsf{s}\left\{\langle [\Delta E_{n_{QM},n_{surr}}^{(\text{elstat})}(\mathbf{r}_n^{QM}, \mathbf{r}_n^{surr})]^2 \rangle_{\mathbf{R}^N} - [\langle \Delta E_{n_{QM},n_{surr}}^{(\text{elstat})}(\mathbf{r}_n^{QM}, \mathbf{r}_n^{surr}) \rangle_{\mathbf{R}^N}]^2\right\} + O(\mathsf{s}^2).$$
(S95)

In this formula, the leading contribution to the electrostatic effective energy $\Delta E_{\text{eff},n_{QM}}^{(\text{elstat})}$ is the ensemble average of the fine-grained electrostatic interaction energy $\Delta E_{n_{QM},n_{surr}}^{(\text{elstat})}$ over the microscopic configurations compatible with the given values of the CG variables $\mathbf{R}^N$, and the lowest-order correction to it is proportional to the dispersion of the electrostatic interaction energy $\Delta E_{i,n_{surr}}^{(\text{elstat})}$. Strictly speaking, Eq. (S95) is incorrect, since typical energies of interaction between the two parts of the system are greater than $k_B T$ at typical temperatures, at least for systems of practical interest, thus the dimensionless parameter of expansion implied by the expansion on the right hand side of Eq. (S95) is not small. However, in a number of physical problems (for example, in the perturbative expansion of the energy of interatomic interaction in a series of inverse powers of the interatomic distance[4,5] or in the WKB approximation[6]) several first terms in mathematically incorrect (e.g., divergent) expansions are known to provide physically reasonable behavior. An answer to the question of whether approximation (S95) is valuable in practice should be partially based on numerical results for various real systems, and we leave this question open in this paper.

## 5. Exact expressions for the contributions to coefficients k from different types of interactions between the QM part and the surroundings

In the approximation of purely **electrostatic** interaction between the QM part and the surroundings, all coefficients $k_{i,j}^{n_{QM},n_{surr}}$ equal zero.

If **induction** interactions between the QM part and the surroundings are included, only those of the coefficients $c_{n_{QM},n_{surr}}^{k,l}$ in Eq. (17) are different from zero that either have a form $c_{n_{QM},n_{surr}}^{k,n_{surr}}$, with $k \neq n_{QM}$, or a form $c_{n_{QM},n_{surr}}^{n_{QM},l}$, with $l \neq n_{surr}$. Then Eq. (22) simplifies to:

$$k_{i,j}^{(\text{ind})n_{QM},n_{surr}} = \begin{cases} \dfrac{-\sum_{k \neq n_{QM}} \left(c_{n_{QM},n_{surr}}^{k,n_{surr}}\right)^2}{1 + \sum_{k \neq n_{QM}} \left(c_{n_{QM},n_{surr}}^{k,n_{surr}}\right)^2 + \sum_{l \neq n_{surr}} \left(c_{n_{QM},n_{surr}}^{n_{QM},l}\right)^2}, & \text{if } i = n_{QM} \text{ \& } j = n_{QM} \\[2ex] \dfrac{c_{n_{QM},n_{surr}}^{j,n_{surr}}}{1 + \sum_{k \neq n_{QM}} \left(c_{n_{QM},n_{surr}}^{k,n_{surr}}\right)^2 + \sum_{l \neq n_{surr}} \left(c_{n_{QM},n_{surr}}^{n_{QM},l}\right)^2}, & \text{if } i = n_{QM} \text{ \& } j \neq n_{QM} \\[2ex] \dfrac{c_{n_{QM},n_{surr}}^{i,n_{surr}}}{1 + \sum_{k \neq n_{QM}} \left(c_{n_{QM},n_{surr}}^{k,n_{surr}}\right)^2 + \sum_{l \neq n_{surr}} \left(c_{n_{QM},n_{surr}}^{n_{QM},l}\right)^2}, & \text{if } i \neq n_{QM} \text{ \& } j = n_{QM} \\[2ex] \dfrac{c_{n_{QM},n_{surr}}^{i,n_{surr}} c_{n_{QM},n_{surr}}^{j,n_{surr}}}{1 + \sum_{k \neq n_{QM}} \left(c_{n_{QM},n_{surr}}^{k,n_{surr}}\right)^2 + \sum_{l \neq n_{surr}} \left(c_{n_{QM},n_{surr}}^{n_{QM},l}\right)^2}, & \text{if } i \neq n_{QM} \text{ \& } j \neq n_{QM} \end{cases} \quad (S96)$$

where the superscript "(ind)" shows that these are the contributions to $k_{i,j}^{n_{QM},n_{surr}}$ stemming from the induction interactions.

Adding **dispersion** interactions changes the expressions for the coefficients $k_{i,j}^{n_{QM},n_{surr}}$ to the following form:

$$k_{i,j}^{(\text{ind+disp})n_{QM},n_{surr}} = \begin{cases} \dfrac{-\sum_{k \neq n_{QM},l} \left(c_{n_{QM},n_{surr}}^{k,l}\right)^2}{1 + \sum_{k,l} \left(c_{n_{QM},n_{surr}}^{k,l}\right)^2}, & \text{if } i = n_{QM} \text{ \& } j = n_{QM} \\[2ex] \dfrac{c_{n_{QM},n_{surr}}^{j,n_{surr}} + \sum_{l} c_{n_{QM},n_{surr}}^{n_{QM},l} c_{n_{QM},n_{surr}}^{j,l}}{1 + \sum_{k,l} \left(c_{n_{QM},n_{surr}}^{k,l}\right)^2}, & \text{if } i = n_{QM} \text{ \& } j \neq n_{QM} \\[2ex] \dfrac{c_{n_{QM},n_{surr}}^{i,n_{surr}} + \sum_{l} c_{n_{QM},n_{surr}}^{n_{QM},l} c_{n_{QM},n_{surr}}^{i,l}}{1 + \sum_{k,l} \left(c_{n_{QM},n_{surr}}^{k,l}\right)^2}, & \text{if } i \neq n_{QM} \text{ \& } j = n_{QM} \\[2ex] \dfrac{\sum_{l} c_{n_{QM},n_{surr}}^{i,l} c_{n_{QM},n_{surr}}^{j,l}}{1 + \sum_{k,l} \left(c_{n_{QM},n_{surr}}^{k,l}\right)^2}, & \text{if } i \neq n_{QM} \text{ \& } j \neq n_{QM} \end{cases} \quad (S97)$$

In all four cases on the right hand side of Eq. (S97), the contribution of the dispersion interactions to $k_{i,j}^{(\text{disp})n_{QM},n_{surr}}$ [defined as the difference between the values of $k_{i,j}^{(\text{ind}+\text{disp})n_{QM},n_{surr}}$ given by Eq. (S97) and the values of $k_{i,j}^{(\text{ind})n_{QM},n_{surr}}$ given by Eq. (S96)] has the order of $O(v^2)$, though the magnitude should in general be smaller in the case of induction contributions since the former include matrix elements of $V_{QM-surr}$ between two different eigenstates of both parts of the system.

Expressions for the contributions to $k_{i,j}^{n_{QM},n_{surr}}$ from the **purely exchange** interactions are technically more involved and provided below in Sec. 10 of the Supplementary Material in the appropriate context. As for the contributions from the exchange-repulsion, exchange-induction and exchange-dispersion interactions, the corresponding analytical expressions would be too involved and for this reason are not presented in this paper.

## 6. Perturbative treatment of the interactions between the QM part and the surroundings

Perturbative calculations of interactions between different parts of molecular systems[7] is a widespread approach in the theory of intermolecular interactions.[8] Though a perturbative treatment may lead to a number of various issues, such as narrow radii of convergence of the series expansions,[9] multiple ways to account for permutation symmetry,[10,11] or divergence at small or large intermolecular distances,[12] these issues have been deeply studied for the interaction of molecules in ground states.[8,11] Applicability of perturbation theories to excited states is less studied, though existing results are promising.[13,14] In the main paper, the perturbative treatment is used as a tool for understanding the structure of interactions between the QM part and the surroundings in the QM/CG-MM model, in a similar way to how perturbative analysis of intermolecular interactions yields the concepts of electrostatic, induction, dispersion, exchange, etc., interactions and serves as a basis for various approximate computational approaches.[8]

Various modifications of the perturbation theory lead to somewhat different expressions for $c_{k,l}^{n_{QM},n_{surr}}$ and $\Delta E_{n_{QM},n_{surr}}$ in terms of $V_{QM-surr}$ (compare to refs. [8,12]). The formulas below are based on one of symmetry-adapted perturbation theories (SAPTs), specifically the symmetrized Rayleigh–Schrödinger (SRS) scheme.[12] The use of other SAPTs seems equally feasible, should one prefer any of them, for example, for the reasons of achieving a better convergence of the power series for a specific molecular system. The SRS scheme is chosen in this paper because it leads to more compact formulas.

In the SRS,[12] the perturbed wave functions and energy levels can be written, in the notations of the main paper and without explicitly showing corrections containing third- and higher-order powers of $V_{QM-surr}$, in the following form

$$|n_{QM} n_{surr}\rangle = \mathcal{A}_{sys} \left[ |n_{QM}^{(0)} n_{surr}^{(0)}\rangle - \sum_{\substack{k \neq n_{QM}, \\ l \neq n_{sys}}} \frac{\langle k^{(0)} l^{(0)} | V_{QM-surr} | n_{QM}^{(0)} n_{surr}^{(0)} \rangle}{E_k^{QM} + E_l^{surr} - E_{n_{QM}}^{QM} - E_{n_{surr}}^{surr}} |k^{(0)} l^{(0)}\rangle \right.$$

$$+ \sum_{\substack{k \neq n_{QM}, \\ l \neq n_{sys}}} \frac{\langle k^{(0)} l^{(0)} | V_{QM-surr} | n_{QM}^{(0)} n_{surr}^{(0)} \rangle \langle n_{QM}^{(0)} n_{surr}^{(0)} | V_{QM-surr} | n_{QM}^{(0)} n_{surr}^{(0)} \rangle}{(E_k^{QM} + E_l^{surr} - E_{n_{QM}}^{QM} - E_{n_{surr}}^{surr})^2} |k^{(0)} l^{(0)}\rangle \quad \text{(S98)}$$

$$\left. - \sum_{\substack{k \neq n_{QM}, \\ l \neq n_{sys}}} \sum_{\substack{k' \neq n_{QM}, \\ l' \neq n_{sys}}} \frac{\langle k^{(0)} l^{(0)} | V_{QM-surr} | k'^{(0)} l'^{(0)} \rangle \langle k'^{(0)} l'^{(0)} | V_{QM-surr} | n_{QM}^{(0)} n_{surr}^{(0)} \rangle}{(E_k^{QM} + E_l^{surr} - E_{n_{QM}}^{QM} - E_{n_{surr}}^{surr})(E_{k'}^{QM} + E_{l'}^{surr} - E_{n_{QM}}^{QM} - E_{n_{surr}}^{surr})} |k^{(0)} l^{(0)}\rangle \right]$$

$$+ O(v^3),$$

$$\Delta E_{n_{QM},n_{surr}} = \frac{\left\langle n_{QM}^{(0)} n_{surr}^{(0)} \middle| V_{QM-surr} \mathcal{A}_{sys} \middle| n_{QM}^{(0)} n_{surr}^{(0)} \right\rangle}{\left\langle n_{QM}^{(0)} n_{surr}^{(0)} \middle| \mathcal{A}_{sys} \middle| n_{QM}^{(0)} n_{surr}^{(0)} \right\rangle} -$$

$$- \frac{1}{\left\langle n_{QM}^{(0)} n_{surr}^{(0)} \middle| \mathcal{A}_{sys} \middle| n_{QM}^{(0)} n_{surr}^{(0)} \right\rangle} \sum_{\substack{k \neq n_{QM} \\ l \neq n_{sys}}} \left[ \frac{\left\langle k^{(0)} l^{(0)} \middle| V_{QM-surr} \middle| n_{QM}^{(0)} n_{surr}^{(0)} \right\rangle}{E_k^{QM} + E_l^{surr} - E_{n_{QM}}^{QM} - E_{n_{surr}}^{surr}} \times \right.$$

$$\left. \times \left( \left\langle n_{QM}^{(0)} n_{surr}^{(0)} \middle| V_{QM-surr} \mathcal{A}_{sys} \middle| k^{(0)} l^{(0)} \right\rangle - \frac{\left\langle n_{QM}^{(0)} n_{surr}^{(0)} \middle| \mathcal{A}_{sys} \middle| k^{(0)} l^{(0)} \right\rangle \left\langle n_{QM}^{(0)} n_{surr}^{(0)} \middle| V_{QM-surr} \mathcal{A}_{sys} \middle| n_{QM}^{(0)} n_{surr}^{(0)} \right\rangle}{\left\langle n_{QM}^{(0)} n_{surr}^{(0)} \middle| \mathcal{A}_{sys} \middle| n_{QM}^{(0)} n_{surr}^{(0)} \right\rangle} \right) \right]$$

$$+ O(v^3), \qquad (S99)$$

where $\left| n_{QM} n_{surr} \right\rangle$, as in the main text, are unnormalized perturbed wave functions for the system, while $\left| n_{QM}^{(0)} n_{surr}^{(0)} \right\rangle$ denotes a simple (that is, not antisymmetrized) products of unperturbed wave functions for the isolated QM part and the surroundings, respectively:

$$\left| n_{QM}^{(0)} n_{surr}^{(0)} \right\rangle = \Psi_{n_{QM}}^{QM} \Psi_{n_{surr}}^{surr}. \qquad (S100)$$

The SRS belongs to the subset of SAPTs that require antisymmetrization of the wave function after building it in the form of power series.[12] The small parameter $v$ is defined by Eq. (32).

Note that if the terms with exchange of electrons between the QM part and the surroundings are neglected, Eqs. (S98) and (S99) reduce to the ordinary Rayleigh–Schrödinger scheme:

$$\begin{aligned}
\left|n_{QM} n_{surr}\right\rangle = &\left[\left|n_{QM}^{(0)} n_{surr}^{(0)}\right\rangle - \sum_{\substack{k \neq n_{QM},\\ l \neq n_{sys}}} \frac{\left\langle k^{(0)} l^{(0)}\left|V_{QM-surr}\right| n_{QM}^{(0)} n_{surr}^{(0)}\right\rangle}{E_k^{QM} + E_l^{surr} - E_{n_{QM}}^{QM} - E_{n_{surr}}^{surr}} \left|k^{(0)} l^{(0)}\right\rangle\right. \\
&+ \sum_{\substack{k \neq n_{QM},\\ l \neq n_{sys}}} \frac{\left\langle k^{(0)} l^{(0)}\left|V_{QM-surr}\right| n_{QM}^{(0)} n_{surr}^{(0)}\right\rangle \left\langle n_{QM}^{(0)} n_{surr}^{(0)}\left|V_{QM-surr}\right| n_{QM}^{(0)} n_{surr}^{(0)}\right\rangle}{(E_k^{QM} + E_l^{surr} - E_{n_{QM}}^{QM} - E_{n_{surr}}^{surr})^2} \left|k^{(0)} l^{(0)}\right\rangle \\
&\left. - \sum_{\substack{k \neq n_{QM}, k' \neq n_{QM},\\ l \neq n_{sys}\ l' \neq n_{sys}}} \frac{\left\langle k^{(0)} l^{(0)}\left|V_{QM-surr}\right| k'^{(0)} l'^{(0)}\right\rangle \left\langle k'^{(0)} l'^{(0)}\left|V_{QM-surr}\right| n_{QM}^{(0)} n_{surr}^{(0)}\right\rangle}{(E_k^{QM} + E_l^{surr} - E_{n_{QM}}^{QM} - E_{n_{surr}}^{surr})(E_{k'}^{QM} + E_{l'}^{surr} - E_{n_{QM}}^{QM} - E_{n_{surr}}^{surr})} \left|k^{(0)} l^{(0)}\right\rangle\right] \\
&+ O(v^3),
\end{aligned}$$
(S101)

$$\Delta E_{n_{QM}, n_{surr}} = \left\langle n_{QM}^{(0)} n_{surr}^{(0)}\left|V_{QM-surr}\right| n_{QM}^{(0)} n_{surr}^{(0)}\right\rangle - \\
- \sum_{\substack{k \neq n_{QM},\\ l \neq n_{sys}}} \frac{\left|\left\langle k^{(0)} l^{(0)}\left|V_{QM-surr}\right| n_{QM}^{(0)} n_{surr}^{(0)}\right\rangle\right|^2}{E_k^{QM} + E_l^{surr} - E_{n_{QM}}^{QM} - E_{n_{surr}}^{surr}} + O(v^3).$$
(S102)

## 7. Accounting for the polarization of the surroundings by the QM part without the use of wave functions of the surroundings

A possible strategy is illustrated below by the analysis of the leading contributions, but this approach can be generalized to include higher-order terms as well. We write the energy of interaction between the two parts $V_{QM\text{-}surr}$ as

$$V_{QM-surr} = \sum_{I \in CG} q_I^{surr} \{^{QM}(\mathbf{R}_I) - \sum_{I \in CG} \left(\boldsymbol{\mu}_I^{surr} \cdot \mathbf{F}^{QM}(\mathbf{R}_I)\right),$$
(S103)

where $I$ enumerates CG particles (typically, molecules) in the surroundings, $q_I^{surr}$ is the charge of the $I$-th CG particle, $\{_{QM}(\mathbf{R}_I)$ is the electrostatic potential created by the electrons and nuclei from the QM part at the position of the $I$-th CG particle $\mathbf{R}_I$, $\boldsymbol{\mu}_I^{surr}$ is the dipole momentum of the $I$-th CG particle, $\mathbf{F}_{QM}(\mathbf{R}_I)$ is the electrostatic field created by the QM part at point $\mathbf{R}_I$, and the terms corresponding to higher order multipoles of the CG particles are ignored. Then the third term on the right hand side of Eq. (31) can be transformed as follows:

$$-\sum_{l \neq n_{surr}} \frac{\left|\langle n_{QM}^{(0)} l^{(0)} | V_{QM-surr} | n_{QM}^{(0)} n_{surr}^{(0)} \rangle\right|^2}{E_l^{surr} - E_{n_{surr}}^{surr}} = -\frac{1}{2} \sum_{\substack{I \in CG \\ I' \in CG}} \sum_{\substack{r=x,y,z \\ s=x,y,z}} \left\{ F_r^{QM,n_{QM}}(\mathbf{R}_I) F_s^{QM,n_{QM}}(\mathbf{R}_I) \times \right. $$
$$\left. \times \sum_{l \neq n_{surr}} \frac{2 \langle l^{(0)} | \tilde{q}_{I,r}^{surr} | n_{surr}^{(0)} \rangle \langle n_{surr}^{(0)} | \tilde{q}_{I',s}^{surr} | l^{(0)} \rangle}{E_l^{surr} - E_{n_{surr}}^{surr}} \right\} \quad \text{(S104)}$$

where $F_r^{QM,n_{QM}}(\mathbf{R}_I)$ is the α-th Cartesian component of $\langle n_{QM}^{(0)} | \mathbf{F}^{QM}(\mathbf{R}_I) | n_{QM}^{(0)} \rangle$, that is the electric field created at point $\mathbf{R}_I$ by the QM part in the quantum state $|n_{QM}^{(0)}\rangle$. In the derivation of Eq. (S104) we used the fact that $\langle l^{(0)} | q_I^{surr} | n_{surr}^{(0)} \rangle = 0$ since $l \neq n_{surr}$. Further simplifications can be made if the following two assumptions are introduced. First, exchange interactions between different CG sites (molecules in the surroundings) can be neglected in the computation of the r.h.s. of Eq. (S104), which implies $|l^{(0)}\rangle = |l_1^{(0)}\rangle |l_2^{(0)}\rangle \ldots |l_N^{(0)}\rangle$, where $|l_I^{(0)}\rangle$ is the unperturbed wave function of the $I$-th CG site (molecule) in the surroundings. Second, if two electronic states of the surroundings $|l^{(0)}\rangle$ and $|n_{surr}^{(0)}\rangle$ differ only in the electronic state of the $I$-th CG site (molecule), $|l_I^{(0)}\rangle$ and $|(n_{surr})_I^{(0)}\rangle$, while all other CG sites (molecules) in the surroundings remain in the same electronic states, then the difference of the energy levels $E_l^{surr} - E_{n_{surr}}^{surr} \approx E_{l_I}^{surr,I} - E_{(n_{surr})_I}^{surr,I}$, where $E_{l_I}^{surr,I}$ is the energy of the $I$-th CG site (molecule) in the electronic state $|l_I^{(0)}\rangle$. With these two assumptions, Eq. (S104) simplifies to

$$-\sum_{l \neq n_{surr}} \frac{\left|\langle n_{QM}^{(0)} l^{(0)} | V_{QM-surr} | n_{QM}^{(0)} n_{surr}^{(0)} \rangle\right|^2}{E_l^{surr} - E_{n_{surr}}^{surr}} = -\frac{1}{2} \sum_{I \in CG} \sum_{\substack{r=x,y,z \\ s=x,y,z}} F_r^{QM,n_{QM}}(\mathbf{R}_I) \Gamma_{rs}^{surr,I,n_{surr}} F_s^{QM,n_{QM}}(\mathbf{R}_I), \quad \text{(S105)}$$

where $r_{rs}^{surr,I,n_{surr}}$ is the polarizability of the *I*-th CG site (molecule) in vacuum in the electronic state $n_{surr}$ defined in the usual way, Eq. (37). If higher order multipoles were included in Eq. (S103), then Eq. (S105) would have additional terms with dipole–quadrupole, quadrupole–quadrupole, etc. polarizabilities coupled to gradients and higher order derivatives of the electric fields $F_r^{QM,n_{QM}}(\mathbf{R}_I)$. Hyperpolarizabilities appear in similar expressions for $O(v^3)$ and higher order in *v* terms in Eq. (31).

Similarly, $\sum_{l \neq n_{surr}} \left(c_{n_{QM},n_{surr}}^{n_{QM},l}\right)^2$ can be approximated, with the use of the same assumptions as those formulated between Eqs. (S104) and (S105), in the following form:

$$\sum_{l \neq n_{surr}} \left(c_{n_{QM},n_{surr}}^{n_{QM},l}\right)^2 = \frac{1}{2} \sum_{I \in CG} \sum_{\substack{r=x,y,z \\ s=x,y,z}} F_r^{QM,n_{QM}}(\mathbf{R}_I) s_{rs}^{surr,I,n_{surr}} F_s^{QM,n_{QM}}(\mathbf{R}_I), \quad (S106)$$

where $s_{rs}^{surr,I,n_{surr}}$ is another characteristic of the *I*-th CG site (molecule) in vacuum in the electronic state $n_{surr}$ defined by Eq. (38) in the main text. Note that the right hand side of Eq. (38) can formally be considered as a partial derivative of the right hand side of Eq. (37) with respect to the energy $E_{(n_{surr})_I}^{surr,I}$. Values of $s_{rs}^{surr,I,n_{surr}}$, similar to polarizabilities, can be taken from reference tables, if available, or otherwise calculated separately for each molecule from the surroundings at a preliminary stage of QM/CG-MM modeling.

Note that in a typical case the resulting eigenvalues and eigenfunctions of the effective QM Hamiltonian are much less sensitive to errors in the description of the polarization of the surroundings by the QM part [Eq. (29) and the second term in Eq. (31)] than those in the polarization of the QM part by the surroundings [Eq. (30) and the third term in Eq. (31)]. In the first and the fourth cases in Eq. (S96), the coefficients $k_{i,j}^{(ind)n_{QM},n_{surr}}$ are on the order of $O(v^2)$, while the contribution of the terms corresponding to the polarization of the surroundings is only on the order of $O(v^4)$; in the second and the third cases, the total values of $k_{i,j}^{(ind)n_{QM},n_{surr}}$ and the contribution of the polarization of the surroundings are on the order of $O(v)$ and $O(v^3)$, respectively. As for the contributions to the interaction energy, the terms arising from the polarization of the QM part are inversely proportional to the energies of electronic excitations in the QM part, while those arising from the polarization of the surroundings are inversely proportional to the energies of excitations in the surroundings. In a typical system studied with a QM/CG-MM model, one may expect to have much easier (in terms of energy) excitations in the QM part than in the surroundings (e.g., much lower excited states in a dye molecule than in a solvent in which it is dissolved), therefore the former contributions to the interaction energy in most cases will be greater than the latter ones. This weaker dependence of the final results on the polarization of the surroundings justifies the use of less accurate assumptions in the derivations of Eqs. (S105) and (S106).

## 8. Analysis of the case with the electrostatic interactions and the polarization of the QM part by the surroundings

Since a wave function of the system for each electronic state in this case can be written as $|n_{QM} n_{surr}\rangle = |n_{QM}\rangle |n_{surr}^{(0)}\rangle$, where $|n_{QM}\rangle$ is the corresponding wave function of the QM part perturbed by the surroundings and $|n_{surr}^{(0)}\rangle$ is the corresponding unperturbed wave function of the surroundings, it is possible to demonstrate that in the basis of $|n_{QM}\rangle$, the matrix for the operator $e^{-s\hat{H}_{eff}^{QM}}$ assumes a diagonal form:

$$\langle i | e^{-s\hat{H}_{eff}^{QM}} | j \rangle = \mathsf{u}_{i,j} CC_{CG} e^{-s V_{CG}^{surr}(\mathbf{R}^N)} e^{-s E_i^{QM}(\mathbf{r}_n^{QM})} \langle e^{-s\Delta E_{i,n_{surr}}(\mathbf{r}_n^{QM},\mathbf{r}_n^{surr})}\rangle_{\mathbf{R}^N}. \qquad (S107)$$

Therefore, the eigenvalues of the effective QM Hamiltonian equal

$$E_{eff,n_{QM}}^{(\text{elstat}+\text{ind QM})} = E_{n_{QM}}^{QM}\left(\mathbf{r}_n^{QM}\right) + V_{CG}^{surr}\left(\mathbf{R}^N\right) + \Delta E_{eff,n_{QM}}^{(\text{elstat})}\left(\mathbf{r}_n^{QM}, \mathbf{R}^N\right) + \Delta E_{eff,n_{QM}}^{(\text{ind QM})}\left(\mathbf{r}_n^{QM}, \mathbf{R}^N\right), \quad (S108)$$

which differs from Eq. (24) by the presence of the induction contribution to the interaction energy:

$$\Delta E_{eff,n_{QM}}^{(\text{ind QM})}\left(\mathbf{r}_n^{QM}, \mathbf{R}^N\right) = -\frac{1}{\text{s}}\left[\ln\langle e^{-\text{s}\Delta E_{n_{QM},n_{surr}}^{(\text{elstat}+\text{ind QM})}(\mathbf{r}_n^{QM},\mathbf{r}_n^{surr})}\rangle_{\mathbf{R}^N} - \ln\langle e^{-\text{s}\Delta E_{n_{QM},n_{surr}}^{(\text{elstat})}(\mathbf{r}_n^{QM},\mathbf{r}_n^{surr})}\rangle_{\mathbf{R}^N}\right], \quad (S109)$$

where

$$\Delta E_{n_{QM},n_{surr}}^{(\text{elstat}+\text{ind QM})} = \Delta E_{n_{QM},n_{surr}}^{(\text{elstat})} - \sum_{k \neq n_{QM}} \frac{\left|\langle k^{(0)}|V_{ext}^{\text{elstat},n_{surr}}|n_{QM}^{(0)}\rangle\right|^2}{E_k^{QM} - E_{n_{QM}}^{QM}} + O(v^3). \quad (S110)$$

Similar to the case discussed in Sec. 4.1 of the Supplementary Material, if the CG variables $\mathbf{R}^N = \{\mathbf{R}_c^N, \mathbf{\Omega}^N\}$ are defined as the centers of mass $\mathbf{R}_c^N$ and the rotation angles $\mathbf{\Omega}^N$ of rigid molecules in the surroundings, then $\Delta E_{n_{QM},n_{surr}}^{(\text{elstat}+\text{ind QM})}$ can be represented as an entity depending on $\mathbf{r}_n^{surr}$ only via the mapping operators. Then Eq. (S109) simplifies to

$$\Delta E_{eff,n_{QM}}^{(\text{ind QM})}\left(\mathbf{r}_n^{QM}, \mathbf{R}^N\right) = -\sum_{k \neq n_{QM}} \frac{\left|\langle k^{(0)}|V_{ext}^{\text{elstat},n_{surr}}(\mathbf{r}_n^{surr}(\mathbf{R}^N))|n_{QM}^{(0)}\rangle\right|^2}{E_k^{QM} - E_{n_{QM}}^{QM}} + O(v^3), \quad (S111)$$

where $V_{ext}^{\text{elstat},n_{surr}}(\mathbf{r}_n^{surr}(\mathbf{R}^N))$ can be expressed in the multipole approximation or based on the distributed multipolar analysis, as discussed in Appendix D.

Alternatively, in the small $\beta$ expansion [see Sec. 4.3 of the Supplementary Material for the discussion of a validity of such an expansion], Eq. (S109) simplifies to

$$\Delta E_{eff,n_{QM}}^{(\text{ind QM})}\left(\mathbf{r}_n^{QM}, \mathbf{R}^N\right) = -\sum_{k \neq n_{QM}} \frac{\left\langle\left|\langle k^{(0)}|V_{ext}^{\text{elstat},n_{surr}}|n_{QM}^{(0)}\rangle\right|^2\right\rangle_{\mathbf{R}^N}}{E_k^{QM} - E_{n_{QM}}^{QM}} + O(v^3) + O(\text{s}). \quad (S112)$$

The wave functions of the QM part $|n_{QM}\rangle$, in contrast to the case of purely electrostatic interactions, do not coincide with the unperturbed wave functions of the QM part $|n_{QM}^{(0)}\rangle$, but they can be directly written in the form of perturbative series, without a need for diagonalization of the matrix for the operator $e^{-s\hat{H}_{eff}^{QM}}$. In this case, QM/CG-MM computations can follow a simpler algorithm described in Scheme 1, with some modifications. The value of $\langle\hat{O}\rangle_{QM/CG-MM}$ should be computed by a modified version of Eq. (27), namely

$$\langle\hat{O}\rangle_{QM/CG-MM} = \frac{\int d\mathbf{r}_n^{QM} d\mathbf{R}^N \sum_{n_{QM}} e^{-s E_{eff,n_{QM}}^{(\text{elstat+ind QM})}(\mathbf{r}_n^{QM},\mathbf{R}^N)} \langle n_{QM}|\hat{O}|n_{QM}\rangle}{\int d\mathbf{r}_n^{QM} d\mathbf{R}^N \sum_{n_{QM}} e^{-s E_{eff,n_{QM}}^{(\text{elstat+ind QM})}(\mathbf{r}_n^{QM},\mathbf{R}^N)}}, \quad (S113)$$

where $E_{eff,n_{QM}}^{(\text{elstat+ind QM})}$ is given by Eq. (S108), and

$$\langle n_{QM}|\hat{O}|n_{QM}\rangle = \langle n_{QM}^{(0)}|\hat{O}|n_{QM}^{(0)}\rangle - \sum_{k\neq n_{QM}} c_{n_{QM},n_{surr}}^{k,n_{surr}} \left(\langle n_{QM}^{(0)}|\hat{O}|k^{(0)}\rangle + \langle k^{(0)}|\hat{O}|n_{QM}^{(0)}\rangle\right) + O(v^2) \quad (S114)$$

(if necessary, with explicit expressions for higher order terms) instead of $\langle n_{QM}^{(0)}|\hat{O}|n_{QM}^{(0)}\rangle$. This simple approach to computing the QM part eigenfunctions $|n_{QM}\rangle$ and matrix elements $\langle n_{QM}|\hat{O}|n_{QM}\rangle$, however, becomes inapplicable as soon as the polarization of the surroundings or dispersion interactions or exchange interactions are accounted for, since each of these leads to a nontrivial projection from the $N_e^{sys}$-particle Hilbert space onto the $N_e^{QM}$-particle Hilbert space. In all those cases, diagonalization of a matrix for $e^{-s\hat{H}_{eff}^{QM}}$ (or an equivalent of this procedure) will be required, as discussed in the main text, Sec. V.B.

## 9. Accounting for the dispersion interactions between the QM part and the surroundings without the use of wave functions of the surroundings

Following the same line of reasoning as in Sec. 7 of the Supplementary Material, we approximate $V_{QM-surr}$ by Eq. (S103), neglect exchange interactions between different molecules in the surroundings, and approximate the differences in the energies of the surrounding by the differences in the energies of separate molecules in the surrounding, $E_l^{surr} - E_{n_{surr}}^{surr} \approx E_{l_I}^{surr,I} - E_{(n_{surr})_I}^{surr,I}$. In this way, we arrive at the following expression for the lowest-order in $v$ contributions of the dispersion interactions to $\Delta E_{n_{QM},n_{surr}}$:

$$\Delta E_{n_{QM},n_{surr}}^{(disp)} = -\sum_{\substack{k \neq n_{QM} \\ l \neq n_{surr}}} \frac{\left|\left\langle k^{(0)} l^{(0)} \left| V_{QM-surr} \right| n_{QM}^{(0)} n_{surr}^{(0)} \right\rangle\right|^2}{E_k^{QM} + E_l^{surr} - E_{n_{QM}}^{QM} - E_{n_{surr}}^{surr}} + O(v^3) =$$

$$-\frac{1}{2} \sum_{k \neq n_{QM}} \sum_{I \in CG} \sum_{\substack{r=x,y,z \\ s=x,y,z}} \sum_{l_I \neq (n_{surr})_I} \left\{ F_r^{QM,k,n_{QM}}(\mathbf{R}_I) F_s^{QM,k,n_{QM}}(\mathbf{R}_I) \times \right. \tag{S115}$$

$$\left. \times \frac{2 \left\langle l_I^{(0)} \left| \tilde{\alpha}_{I,r}^{surr} \right| (n_{surr})_I^{(0)} \right\rangle \left\langle (n_{surr})_I^{(0)} \left| \tilde{\alpha}_{I,s}^{surr} \right| l_I^{(0)} \right\rangle}{E_{l_I}^{surr,I} - E_{(n_{surr})_I}^{surr,I}} \cdot \frac{1}{1 + \frac{E_{l_I}^{surr,I} - E_{(n_{surr})_I}^{surr,I}}{E_k^{QM} - E_{n_{QM}}^{QM}}} \right\} + O(v^3)$$

where

$$F_r^{QM,k,n_{QM}}(\mathbf{R}_I) = \left\langle k^{(0)} \left| F_r^{QM}(\mathbf{R}_I) \right| n_{QM}^{(0)} \right\rangle \tag{S116}$$

and the other notations are the same as in Sec. V.B of the main text and 7 of the Supplementary Material. If the gaps between different energy levels in the surroundings are greater than those in the QM part, which can be considered as a typical situation in a system modeled with a QM/CG-MM approach (otherwise one would probably use a finer description for the surroundings and would not need to use QM description for the active part), then Eq. (S115) simplifies to

$$\Delta E_{n_{QM},n_{surr}}^{(\text{disp})} = -\frac{1}{2} \sum_{k \neq n_{QM}} \sum_{I \in CG} \sum_{\substack{r=x,y,z \\ s=x,y,z}} F_r^{QM,k,n_{QM}}(\mathbf{R}_I) \Gamma_{rs}^{surr,I,n_{surr}} F_s^{QM,k,n_{QM}}(\mathbf{R}_I) + O(v^3), \quad (S117)$$

which differs from the expression for the induction energy coming from the polarization of the surroundings by the QM part, Eq. (S105), only by the use of off-diagonal elements of the matrix for the electrostatic field, Eq. (46), and extra summation over the corresponding intermediate states enumerated by the index $k$. Then the interaction energy can be written as

$$\Delta E_{n_{QM},n_{surr}}^{(\text{elstat}+\text{ind}+\text{disp})} = \Delta E_{n_{QM},n_{surr}}^{(\text{elstat})} + \Delta E_{n_{QM},n_{surr}}^{(\text{ind})}$$
$$- \frac{1}{2} \sum_{k \neq n_{QM}} \sum_{I \in CG} \sum_{\substack{r=x,y,z \\ s=x,y,z}} F_r^{QM,k,n_{QM}}(\mathbf{R}_I) \Gamma_{rs}^{surr,I,n_{surr}} F_s^{QM,k,n_{QM}}(\mathbf{R}_I) + O(v^3). \quad (S118)$$

The technique of representing the dispersion energy of intermolecular interaction via an integral of a product of polarizabilities at imaginary frequencies[8,15] does not seem applicable to the right hand side of Eq. (S115) in a general case, since this technique requires that $E_k^{QM} - E_{n_{QM}}^{QM} > 0$ and $E_l^{surr} - E_{n_{surr}}^{surr} > 0$, which is not necessarily true for arbitrary $n_{QM}$ and $n_{surr}$.

As for the computation of the coefficients $k_{i,j}^{(\text{ind}+\text{disp})n_{QM},n_{surr}}$ by Eq. (S97), in addition to the values computed in Sec. 7 of the Supplementary Material, only $\sum_{l \neq n_{surr}} c_{n_{QM},n_{surr}}^{i,l} c_{n_{QM},n_{surr}}^{j,l}$ are required. Applying the same approximations as those used to derive Eq. (44), we arrive at the following result:

$$\sum_{l \neq n_{surr}} c_{n_{QM},n_{surr}}^{i,l} c_{n_{QM},n_{surr}}^{j,l} = \frac{1}{2} \sum_{I \in CG} \sum_{\substack{r=x,y,z \\ s=x,y,z}} F_r^{QM,i,n_{QM}}(\mathbf{R}_I) S_{rs}^{surr,I,n_{surr}} F_s^{QM,j,n_{QM}}(\mathbf{R}_I) + O(v^3). \quad (S119)$$

## 10. Analysis of the purely exchange interactions between the QM part and the surroundings

As follows from plugging $|n_{QM} n_{surr}\rangle = \mathcal{A}_{sys}(|n_{QM}^{(0)}\rangle |n_{surr}^{(0)}\rangle)$ into Eq. (22),

$$k_{i,j}^{(pur.exch.)\, n_{QM}, n_{surr}} =$$

$$= \frac{\int d\mathbf{r} d\mathbf{r}' d\mathbf{r}^{surr}\, \Psi_i^{QM}(\mathbf{r}) \mathcal{A}_{sys}\left(\Psi_{n_{QM}}^{QM}(\mathbf{r}) \Psi_{n_{surr}}^{surr}(\mathbf{r}^{surr})\right) \mathcal{A}_{sys}\left(\Psi_{n_{QM}}^{QM}(\mathbf{r}') \Psi_{n_{surr}}^{surr}(\mathbf{r}^{surr})\right) \Psi_j^{QM}(\mathbf{r}')}{\int d\mathbf{r} d\mathbf{r}^{surr}\, \mathcal{A}_{sys}\left(\Psi_{n_{QM}}^{QM}(\mathbf{r}) \Psi_{n_{surr}}^{surr}(\mathbf{r}^{surr})\right) \mathcal{A}_{sys}\left(\Psi_{n_{QM}}^{QM}(\mathbf{r}) \Psi_{n_{surr}}^{surr}(\mathbf{r}^{surr})\right)}$$

$$- \mathsf{u}_{i, n_{QM}} \mathsf{u}_{j, n_{QM}}. \quad (S120)$$

By using the following representation of the total antisymmetrizer

$$\mathcal{A}_{sys} = \frac{(N_e^{QM}!)(N_e^{surr}!)}{N_e^{sys}!}\left(1 - P^{(1)} + P^{(>1)}\right) \mathcal{A}_{QM} \mathcal{A}_{surr}, \quad (S121)$$

where $P^{(1)}$ is the sum of all possible pairwise permutations of electrons between the two parts of the system

$$P^{(1)} = \sum_{i_{QM}=1}^{N_e^{QM}} \sum_{i_{surr}=1}^{N_e^{surr}} P_{i_{QM} \leftrightarrow i_{surr}}, \quad (S122)$$

$P^{(>1)}$ is the sum of certain simultaneous permutations of at least two pairs of electrons between the two parts of the system, and $P_{i_{QM} \leftrightarrow i_{surr}}$ is the operator permuting the $i_{QM}$-th electron in the QM part and the $i_{surr}$-th electron in the surroundings, we arrive at the following expression for the purely exchange contributions to the coefficients $k_{i,j}^{n_{QM}, n_{surr}}$:

$$k_{i,j}^{(pur.exch.)\,n_{QM},n_{surr}} = \begin{cases} \left(\dfrac{1}{C_{N_e^{sys}}^{N_e^{QM}}}-1\right) - \dfrac{N_e^{QM} N_e^{surr}}{C_{N_e^{sys}}^{N_e^{QM}}} \Bigg[\int d\mathbf{x}d\mathbf{x}' \ldots_{n_{surr}}^{surr}(\mathbf{x},\mathbf{x}') \times \\ \quad \times \left(\tilde{\ldots}_{i,n_{QM}}^{QM}(\mathbf{x},\mathbf{x}') + \tilde{\ldots}_{j,n_{QM}}^{QM}(\mathbf{x},\mathbf{x}') - \ldots_{n_{QM}}^{QM}(\mathbf{x},\mathbf{x}')\right)\Bigg] + O(s^3), & \text{if } i=n_{QM} \ \& \ j=n_{QM}; \\[1em] -\dfrac{N_e^{QM} N_e^{surr}}{C_{N_e^{sys}}^{N_e^{QM}}}\Bigg[\int d\mathbf{x}d\mathbf{x}' \ldots_{n_{surr}}^{surr}(\mathbf{x},\mathbf{x}')\tilde{\ldots}_{j,n_{QM}}^{QM}(\mathbf{x},\mathbf{x}')\Bigg] + O(s^3), & \text{if } i=n_{QM} \ \& \ j\neq n_{QM}; \\[1em] -\dfrac{N_e^{QM} N_e^{surr}}{C_{N_e^{sys}}^{N_e^{QM}}}\Bigg[\int d\mathbf{x}d\mathbf{x}' \ldots_{n_{surr}}^{surr}(\mathbf{x},\mathbf{x}')\tilde{\ldots}_{i,n_{QM}}^{QM}(\mathbf{x},\mathbf{x}')\Bigg] + O(s^3), & \text{if } i\neq n_{QM} \ \& \ j=n_{QM}; \\[1em] O(s^3), & \text{if } i\neq n_{QM} \ \& \ j\neq n_{QM}; \end{cases}$$

(S123)

where $C_{N_e^{sys}}^{N_e^{QM}} = N_e^{sys}!/(N_e^{QM}!N_e^{surr}!)$ is a binomial coefficient, , $\ldots_{n_{surr}}^{surr}(\mathbf{x},\mathbf{x}')$ is the one-particle off-diagonal density matrix for the surroundings:

$$\ldots_{n_{surr}}^{surr}(\mathbf{x},\mathbf{x}') = \int d\mathbf{r}_2^{surr} d\mathbf{r}_3^{surr} \cdots d\mathbf{r}_{N_e^{surr}}^{surr} \ \Psi_{n_{surr}}^{surr}\left(\mathbf{x},\mathbf{r}_2^{surr},\mathbf{r}_3^{surr},\cdots,\mathbf{r}_{N_e^{surr}}^{surr}\right)\Psi_{n_{surr}}^{surr}\left(\mathbf{x}',\mathbf{r}_2^{surr},\mathbf{r}_3^{surr},\cdots,\mathbf{r}_{N_e^{surr}}^{surr}\right),$$

(S124)

$\ldots_{n_{QM}}^{QM}(\mathbf{x},\mathbf{x}')$ is the similarly defined one-particle off-diagonal density matrix for the QM part, $\tilde{\ldots}_{i,n_{QM}}^{QM}(\mathbf{x},\mathbf{x}')$ is defined as

$$\tilde{\ldots}_{i,n_{QM}}^{QM}(\mathbf{x},\mathbf{x}') = \int d\mathbf{r}_2^{QM} d\mathbf{r}_3^{QM} \cdots d\mathbf{r}_{N_e^{QM}}^{QM} \ \Psi_i^{QM}\left(\mathbf{x},\mathbf{r}_2^{QM},\mathbf{r}_3^{QM},\cdots,\mathbf{r}_{N_e^{QM}}^{QM}\right)\Psi_{n_{QM}}^{QM}\left(\mathbf{x}',\mathbf{r}_2^{QM},\mathbf{r}_3^{QM},\cdots,\mathbf{r}_{N_e^{surr}}^{QM}\right),$$

(S125)

and **x** and **x'** are three-dimensional vectors representing coordinates of an electron not integrated out in the one-particle density matrices, Eqs. (S124) and (S125).

The $O(s^3)$ terms in Eq. (S123) stand for the expressions that contain the operator $P^{(>1)}$. Explicitly, these terms involve integrals of products of three or more off-diagonal density matrices and, therefore, decay faster with the growth of distances between atoms in two different parts of the system than the terms shown explicitly in Eq. (S123). The notation $O(s^3)$ is motivated by the form that the corresponding integrals assume in the Hartree-Fock approximation. The integrals shown explicitly in Eq. (S123) simplify to

$$N_e^{QM} N_e^{surr} \int d\mathbf{x} d\mathbf{x}' \ldots_{n_{surr}}^{surr}(\mathbf{x},\mathbf{x}') \tilde{\ldots}_{i,n_{QM}}^{QM}(\mathbf{x},\mathbf{x}') = \sum_{p=1}^{N_e^{QM}} \sum_{q=1}^{N_e^{surr}} s_{i_p,(n_{surr})_q} s_{(n_{QM})_p,(n_{surr})_q}, \quad \text{(S126)}$$

$$N_e^{QM} N_e^{surr} \int d\mathbf{x} d\mathbf{x}' \ldots_{n_{surr}}^{surr}(\mathbf{x},\mathbf{x}') \ldots_{n_{QM}}^{QM}(\mathbf{x},\mathbf{x}') = \sum_{p=1}^{N_e^{QM}} \sum_{q=1}^{N_e^{surr}} s^2_{(n_{QM})_p,(n_{surr})_q}, \quad \text{(S127)}$$

where $s_{(n_{QM})_p,(n_{surr})_q}$ is the overlap integral between the $p$-th unperturbed molecular spin-orbital $\{_{(n_{QM})_p}^{QM}(\mathbf{x})\}$ for the QM part occupied in the quantum state $n_{QM}$ and the $q$-th unperturbed molecular spin-orbital $\{_{(n_{surr})_q}^{surr}(\mathbf{x})\}$ for the surroundings occupied in the quantum state $n_{surr}$:

$$s_{(n_{QM})_p,(n_{surr})_q} = \int d\mathbf{x} \{_{(n_{QM})_p}^{QM}(\mathbf{x}) \{_{(n_{surr})_q}^{surr}(\mathbf{x}). \quad \text{(S128)}$$

Hence, the integrals explicitly shown in Eq. (S123) are all of the order $O(s^2)$, where $s$ is a typical or maximal value of the overlap integrals $s_{(n_{QM})_p,(n_{surr})_q}$. As for the integrals omitted from Eq. (S123), they are of the order $O(s^3)$ or higher in terms of the overlap integrals between the two different parts of the system. As is known from the theory of intermolecular interactions, such overlap integrals exponentially decay with the shortest distance between the atoms in the QM part and the atoms in the surroundings.

In practice, computation of the eigenvalues and eigenfunctions of the effective QM Hamiltonian may follow different strategies. One way may be to approximate the off-diagonal density matrix of the surroundings $\ldots_{n_{surr}}^{surr}(\mathbf{x},\mathbf{x}')$ as a sum of density matrices $\ldots_{n_{surr}}^{surr,I}(\mathbf{x},\mathbf{x}')$ of the molecules constituting the surroundings

$$\ldots_{n_{surr}}^{surr}(\mathbf{x},\mathbf{x}') \simeq \sum_{I \in surr} \ldots_{n_{surr}}^{surr,I}(\mathbf{x},\mathbf{x}'), \quad (S129)$$

and then to use the resulting functions $\ldots_{n_{surr}}^{surr}(\mathbf{x},\mathbf{x}')$ to compute $k_{i,j}^{(pur.exch.)n_{QM},n_{surr}}$ and, subsequently, the matrix elements of the effective Hamiltonian. The approximation, Eq. (S129), is motivated by the fact that the main contributions to the integrals in Eq. (S123) come from the regions of close intermolecular contacts between the QM part and the surroundings, which are typically not entangled with each other due to spatial separation. The required values of $\ldots_{n_{surr}}^{surr,I}(\mathbf{x},\mathbf{x}')$ can be obtained from preliminary QM computations for separate molecules in the surroundings, similarly to polarizabilities.

Another practical way may be to choose an empirical approximation for $\Delta E_{eff,i}^{(pur.exch.)}$ directly, without computing the values of coefficients $k_{i,j}^{(pur.exch.)n_{QM},n_{surr}}$. For example, one may follow an analogy with the treatment of the energy of charge penetration in the theory of intermolecular interactions,[16] and compute the contribution to the eigenvalues of the effective Hamiltonian in the following way:

$$\Delta E_{eff,i}^{(pur.exch.)}\left(\mathbf{r}_n^{QM},\mathbf{R}^N\right) \simeq \frac{\ln C_{N_e^{sys}}^{N_e^{QM}}}{S} + \frac{1}{S} \langle \sum_{j \in QM} \sum_{k \in surr} A_{ijk} e^{-\Gamma_{ijk} r_{jk}} \rangle_{\mathbf{R}^N}, \quad (S130)$$

where $j$ and $k$ enumerate atoms in the QM part and the surroundings, respectively, $r_{jk}$ is the distance between the $j$-th and the $k$-th atoms, and the coefficients $A_{ijk}$ and $\alpha_{ijk}$ are to be found by calibration against full QM computations for smaller systems. Other functional forms with fitted parameters could also be suggested by analogy with the existing approximations for the Pauli repulsion energy.[17]

# REFERENCES (SUPPORTING INFORMATION)